\documentclass[11pt]{article}
\usepackage[utf8]{inputenc}

\usepackage{amsthm, amsfonts, amssymb, amsmath,enumitem, natbib, bbm, mathabx}
\usepackage{cases, amsthm}
\usepackage{fullpage}
\usepackage{mathtools}
\usepackage{mathrsfs}
\usepackage{wasysym}
\usepackage{verbatim}
\usepackage{fancyhdr}
\usepackage{graphicx}
\usepackage{bbm}
\usepackage{caption}
\usepackage{subcaption}
\usepackage{authblk}
\usepackage{setspace}
\usepackage{booktabs}
\usepackage{float}
\usepackage{array}
\usepackage[linesnumbered, vlined, longend, ruled]{algorithm2e} 
\usepackage[colorlinks=true, pdfstartview=FitV,
linkcolor=blue, citecolor=blue, urlcolor=blue]{hyperref}
\usepackage{cleveref}

\newtheorem{definition}{Definition}[section]
\newtheorem{theorem}[definition]{Theorem}

\newtheorem{lemma}[definition]{Lemma}

\newtheorem{remark}[definition]{Remark}
\newtheorem{assumption}[definition]{Assumption}

\renewcommand{\hat}{\widehat}

\theoremstyle{remark}

\usepackage[top=1in, bottom=1in, left=1in, right=1in]{geometry}

\title{Doubly Robust and Efficient Calibration of Prediction Sets for Right-Censored Time-to-Event Outcomes}

\author[1]{Rebecca Farina}
\author[2]{Eric J. Tchetgen Tchetgen}
\author[1]{Arun Kumar Kuchibhotla}

\affil[1]{Department of Statistics \& Data Science, Carnegie Mellon University}
\affil[2]{Department of Statistics \& Data Science, University of Pennsylvania}

\date{}

\begin{document}

\maketitle

\begin{abstract}
Our objective is to construct well-calibrated prediction sets for a time-to-event outcome subject to right-censoring with guaranteed coverage. Inspired by modern conformal inference, our approach avoids the need for a well-specified parametric or semiparametric survival model. Unlike existing conformal methods for survival data, which assume Type-I censoring with fully observed censoring times, we consider the more common right-censoring setting in which only the censoring time or only the event time is observed, whichever comes first. 
Under a standard conditional independence censoring condition, we propose and analyze several lower prediction bounds for the survival time of a future observation, including  inverse-probability-of-censoring weighting, and its augmented version based on the semiparametric efficient influence function for the relevant marginal quantile of the outcome  accounting for dependent censoring.
We formally establish asymptotic coverage guarantees of the proposed methods, and demonstrate both theoretically and through empirical experiments, that the augmented approach substantially improves efficiency over all other proposed methods. Specifically, its coverage error bound is doubly robust, and therefore of second order, thus ensuring that it is asymptotically negligible relative to the coverage error of the other methods.
\end{abstract}

\section{Introduction}\label{sec:intro}
\subsection{Background}
A fundamental challenge of survival analysis is the potential for a person's event time to not be completely observed if, either she were to drop out, or the study ended before she experienced the event of primary interest: a phenomenon commonly known as \emph{right-censoring} (see, e.g., \cite{kleinbaum1996survival, hosmer2008applied, collett2023modelling}). While classical tools such as the Kaplan–Meier estimator of the marginal survival curve \citep{Kaplan}, and the Cox proportional hazards regression model \citep{Cox} are well established, there remains a lack of flexible frameworks that provide rigorous uncertainty quantification for predictive inference of right-censored time-to-event outcomes.

Conformal predictive inference, introduced by \cite{Vovk}, provides a general blueprint for constructing well-calibrated prediction intervals, with marginal coverage guarantees under relatively mild conditions. The main advantage of conformal prediction is that it provides valid finite-sample prediction intervals for a new observation exchangeable with the training sample, without imposing strong distributional assumptions on the underlying data generating mechanism. Conformal prediction methods have quickly developed for a wide variety of settings that depart from the canonical i.i.d setting, including covariate shift \citep{tibshirani2019conformal, yang2024doubly}, missing covariates \citep{zaffran2024predictive}, time series analysis \citep{stankeviciute2021conformal}, high-dimensional regression \citep{chen2016trimmed} to name a few. However, conformal predictive inference for censored time-to-event outcomes remains under-developed. 

In this paper, we propose several new methods to construct well-calibrated and efficient prediction bounds for survival times under standard right-censoring. Notably, to calibrate our prediction intervals, we propose two separate approaches based on inverse-propability-of-censoring. \citep{robins1993information, robins1992recovery, robins2000correcting}. The first approach (a) constructs a consistent inverse-probability-of-censoring weighting (IPCW) estimator of the marginal probability that the failure times exceed an estimated covariate-dependent lower bound, while the second approach (b) is based on an augmented-inverse-probability-of-censoring weighting (AIPCW) estimator for the same marginal probability. Although IPCW and AIPCW methods are well developed for estimation and inference in standard survival analysis settings, (a) and (b) appear to provide their first use for obtaining well-calibrated prediction intervals. In addition, we consider two so-called outcome regression-based methods that obviate the need for estimating the censoring mechanism: the Outcome Regression method (OR), which entails estimating the conditional quantile function for the outcome at the desired nominal level; and the Calibrated Outcome Regression method (COR) which adjusts the quantile level to match the desired nominal marginal coverage.

\subsection{Problem setup}\label{sec:intro_probsetup}
Let $X\in\mathbb{R}^d$ denote a unit's vector of covariates, $T\in\mathbb{R}_+$ the unit's  survival time and $C\in\mathbb{R}_+$  the unit's potential censoring time. We denote by $P_{(X,T)}$ the joint distribution of $(X,T)$, and by $P_X$, $P_T$, $P_{T\mid X}$, etc., the corresponding marginal and conditional distributions. While $X$ is fully observed, the survival time $T$ is right-censored in the sense that only $T$ or $C$ is observed, whichever comes first, that is, $Y = \min\{T,C\}$ is observed, together with the censoring indicator $\Delta=\mathbf{1}\{T\leq C\}$. Thus, the fully observed data is given by the i.i.d. sample $\mathcal{D}=(X_i,Y_i,\Delta_i)_{i=1}^n$. We denote the underlying (unobserved) full data by $\mathcal{Z}=(X_i,T_i)_{i=1}^n$.

We now introduce additional notation used throughout. Given a generic function $f$, whenever necessary, we denote as $f^\star$ the true unknown function which generated the sample and by $\hat{f}$ a corresponding estimator obtained from the observed data. Let $S_{T\mid X}(\cdot\mid X)$ denote the survival function of the time-to-event $T$ given $X$, i.e., $S_{T\mid X}(t\mid X) = \mathbb{P}(T > t\mid X)$, and $S_{C\mid X}(\cdot\mid X)$ the survival function of $C$ given $X$. For any $\gamma\in(0,1)$, the conditional quantile function of $T \mid X$ is defined as $q_{T\mid X}(\gamma\mid X)=\inf\{t\in\mathbb{R}_+:S_{T\mid X}(t \mid X) \le 1-\gamma\}$.

We aim to obtain predictive bounds $\widehat{L}(X)$ and $\widehat{U}(X)$ for time-to-event outcome $T$ leveraging the covariate vector $X$ with a marginal coverage guarantee, i.e., $\mathbb{P}(\widehat{L}(X) \le T \le \widehat{U}(X)) \ge 1 - \alpha$.
For comparison with most of the existing literature, we focus on lower predictive bounds (LPBs), even though all our methods can provide two-sided predictive bounds; see~\eqref{eq:IPCW} and subsequent discussion. 
\begin{definition}
     Let $\hat L(\cdot)$ be a function estimated from the observed data $\mathcal{D}$ and $\alpha\in[0,1]$ be a non-stochastic level. $\hat L$ is a marginally calibrated LPB for $T$ at level $\alpha$ if
     \begin{equation*}
    \mathbb{P}(T\ge\hat{L}(X)) \ge 1-\alpha,  
    \end{equation*}
    where the probability is taken with respect to a new data point~$(X,T)\sim P_{(X,T)}$ and the observed data $\mathcal{D}$. $\hat{L}(\cdot)$ is an asymptotically marginally calibrated LPB for $T$ at level $\alpha$ if
    \[
    \mathbb{P}(T \ge \hat{L}(X)) \ge 1 - \alpha - o(1) \quad\mbox{as}\quad n\to\infty.
    \]
\end{definition}
Given the distribution of $(X, T)$, there are several possibilities for an oracle LPB. For instance, one can take $L(\cdot)$ to be the $\alpha$-th quantile of the marginal distribution $T$, which is a constant independent of $X$. Observing that a coverage guarantee conditional on the covariate vector $X$ would be much stronger and more broadly applicable, we consider the oracle LPB as the $\alpha$-th conditional quantile of $T$ given $X$, i.e., $q_{T\mid X}^\star(\alpha\mid \cdot)$. 

Given that we often have one observed dataset, one might require coverage conditional on the observed data. 
In this case, the probably-approximately-correct (PAC) type LPB can provide a more relevant goal for the coverage guarantee.
\begin{definition}\label{def:PAC}
     Let $\hat L(\cdot)$ be a function estimated from the observed data $\mathcal{D}$ and $\alpha,\epsilon\in[0,1]$ be fixed levels. $\hat L$ is an $(\alpha,\epsilon)$-probably approximately correct (PAC) LPB for $T$ if
     \begin{equation*}
    \mathbb{P}\left(\mathbb{P}(T\ge\hat{L}(X) \mid \mathcal{D}) \ge 1-\alpha\right) \ge 1 - \epsilon,  
    \end{equation*}
    where the probability inside is calculated with respect to the new data point $(X,T)\sim P_{(X,T)}$ independent of $\mathcal{D}$. $\hat{L}(\cdot)$ is an asymptotic $(\alpha,\epsilon)$-PAC LPB for $T$ if
    \[
    \mathbb{P}\left(\mathbb{P}(T \ge \hat{L}(X)|\mathcal{D}) \ge 1 - \alpha - o_p(1)\right) \ge 1 - \epsilon \quad\mbox{as}\quad n\to\infty,
    \]
    where the $o_p(1)$ random variable can depend on $\epsilon$. 
\end{definition}
As noted in \citet{Candès}, by definition, $Y$ cannot be larger than $T$, thus any valid LPB for $Y$ is trivially an LPB for $T$. Nevertheless, a valid lower bound for $Y$ may yield an overly conservative lower bound for $T$; for this reason, in this work, we focus on constructing bounds directly for the underlying survival time $T$ of interest, rather than the observed censored time $Y$.

\subsection{Related work}
Recent conformal prediction methods for survival data include Cox regression-based  \citep{teng2021t} and random survival forests-based \citep{bostrom2017conformal, bostrom2019predicting}  methods; as well as \cite{Candès} and \cite{Gui}, who propose to construct covariate-dependent lower predictive bounds, however, assuming that censoring times are observed on all units in training and validation samples, also known as \emph{Type I Censoring}. Formally, the observed data consists of the tuples $(X, C, Y)$. Under this setting, \citet{Candès} established that any distribution-free well-calibrated LPB for $T$ must necessarily be an LPB for $Y$. This motivated their proposed approach, which entails discarding any observation with early censoring time, below a user-specified constant $c_0$, combined with weighted conformal inference \citep{tibshirani2019conformal} to correct for the distribution shift introduced by the initial filtering step. Based on this approach, \citet{Gui} incorporate a covariate-dependent cutoff, making the approach substantially less conservative. Under Type I censoring, their method yields an LPB, which is asymptotically PAC-calibrated in the sense of Definition~\ref{def:PAC}, with the $o_p(1)$ term being the maximum of the estimation error of the conditional survival function of the censoring time given $X$ and an $\left(\log(1/\epsilon)/n\right)^{1/2}$-order term; see Theorem 3 of~\cite{Gui}.

By requiring that censoring times are fully observed, \citet{Candès} and \citet{Gui} methods are primarily useful in applications where potential censoring times are determined at the start of a unit's follow-up, such as in experimental settings where administrative censoring times are pre-determined, while other common forms of censoring, such as loss-to-follow-up and other related protocol departures that can seldom if ever be prevented by design. % Censoring due to dropout or protocol violation can seldom be prevented, even in well-controlled environments, such as randomized clinical trials, therefore excluding the possible utility of these prior methods from the most common form of censoring encountered in practice. Furthermore, 
Another apparent limitation of \citet{Candès} and \citet{Gui} methods is their over-reliance on a cutoff tuning parameter that induces artificial censoring of units with censoring time in the left tail of their distribution, possibly censoring units with observed failure times. This form of artificial censoring introduces a bias-variance trade-off that may be challenging to balance in practice. 
Our proposed methods target an asymptotic PAC guarantee analogous to theirs. However, as detailed in the next Section, we achieve this goal while (i) accommodating a more common form of right-censoring; (ii) obviating the need for artificial censoring or a corresponding tuning parameter; and (iii) obtaining a doubly robust coverage error guarantee,  in the sense that our coverage error slack is not only of second order, but actually of the mixed-bias product form of the estimation bias of the censoring and failure time laws given covariates, plus an irreducible root-$n$ error term.

Recent works have considered alternative strategies for obtaining well-calibrated predictions under conventional censoring. 
In the Type-I censoring setting, \citet{wan2025distributionfree} propose distribution-free lower predictive bounds by combining hybrid quantile learning with a data-filtered, adaptive conformal calibration scheme. 
Moving beyond Type-I censoring, \citet{sesia2025doubly} extend the conformalized survival analysis framework of \citet{Candès} and \citet{Gui} to general right-censoring by first imputing unobserved censoring times under conditionally independent censoring and then applying weighted conformal inference to the imputed data. %\citep{sesia2025doubly} thus targets one-sided lower predictive bounds for survival times, but relies on an explicit imputation step and does not pursue semiparametrically efficient calibration of the lower bound. 
A recent work by \citet{qin2025conformal} proposes a bootstrap resampling approach to build conformal predictive intervals for general right-censored data, achieving asymptotic marginal validity under conditionally independent censoring.
Building on our work, \citet{si2025training} study lower prediction bounds with training-set conditional APAC-type guarantees, meaning that the approximation is placed on the confidence level, rather than on the miscoverage probability as in our asymptotic PAC formulation,  while conditioning on the training sample; they adopt the efficient influence function correction introduced by our approach. %modifying the target guarantee from our PAC-type formulation to a training-set conditional APAC formulation. 
Moreover, \citet{yi2025survival} construct conformal prediction intervals employing inverse censoring weighting and quantile regression as a flexible base learner. A different line of work focuses on survival curves rather than on the time-to-event: \citet{sesia2025conformal} construct conformal bands for individual survival curves for screening-type tasks by combining inverse censoring weighting with multiple-testing FDR control. We also note the work of  \citet{park2025ksp}, which performs post-hoc calibration of predicted survival curves by reducing a Kolmogorov-Smirnov discrepancy between predicted and empirical survival probabilities.
Finally, \citet{meixide} introduced a prediction inference framework for interval-censored data, an important but different focus to ours. While these papers substantially advance conformal uncertainty quantification for censored outcomes, they do not simultaneously meet our specific desiderata (i)--(iii): in particular, for the setting and goal we consider, we provide a doubly robust procedure which attains semiparametric efficiency (under standard regularity conditions) and yields second-order coverage error control.

In a separate research strand, inverse-probability-of-censoring weighting (IPCW) and augmented-inverse-probability-of-censoring weighting (AIPCW) have played a pivotal role in the development of robust and efficient methodologies for censored data; primarily in the context of inference for a finite dimensional functional of interest.  The foundational semiparametric estimation robustness and efficiency theory in the presence of censoring, based on IPCW and AIPCW, was pioneered by Robins and colleagues \citep{robins1992recovery, robins1993information}.
See \cite{laan2003unified, tsiatis2007semiparametric} for an introductory textbook overview of modern semiparametric efficiency theory for censored and other missing data problems. Our work extends the recent paper of \citet{yang2024doubly} which to the best of our knowledge provides the first application of semiparametric efficiency theory to a conformal prediction setting subject to covariate shift. While their proposed approach can equivalently be viewed as an outcome missing at random, their proposed methods do not directly apply to the censored survival outcome setting which presents a number of new challenges we address in this work. 

\subsection{Our contributions}
Motivated by the limitations of existing methods, we develop a new approach for well-calibrated predictive inference for standard right-censored survival outcomes, whereby either only the censoring or the primary event time is observed, whichever comes first. 
Our goal is to derive predictive bounds while avoiding unnecessary modeling assumptions. The first of our two proposed methods incorporates IPCW, which data adaptively accounts for the unknown censoring mechanism, provided that measured covariates are sufficiently rich to explain any dependence between censoring and the event time of primary interest.  This completely obviates the need for artificial censoring and the required tuning of a cutoff parameter, while recovering robust (asymptotically) well-calibrated predictive bounds for survival times. 

A well-known limitation of IPCW is that it is potentially inefficient due to discarding information contained in censored observations. Thus, in addition, we introduce an augmented IPCW approach, which recovers information from censored observations with the potential for significant efficiency improvements over IPCW. Using AIPCW also leads to a desirable double robustness property, which ensures that the error in the coverage bound of the AIPCW prediction sets, due to estimation of nuisance parameters (mainly  of the conditional survival curve for the primary outcome and for the censoring mechanism given covariates, respectively), is of mixed bias product form \citep{rotnitzky2021characterization}. This ensures that our method can provide valid bounds even if one of the required survival curves is misspecified,  offering enhanced flexibility and reliability in practical applications. 

Importantly, the proposed methodology for deriving predictive bounds does not depend on the specific choice of a non-conformity score. The resulting bounds preserve their validity, and the AIPCW-based approach remains doubly robust, regardless of the choice of non-conformity score. This flexibility highlights the generality and robustness of our proposed framework. Furthermore, if the non-conformity score is properly chosen~\eqref{eq:conditional-quantile-non-conformity-score}, to match with the intuition of conditional quantile as the oracle LPB, then our methods also recover approximate conditional coverage; see Theorem 4 of~\cite{Gui} for a related result.

As a complementary contribution, we introduce two simple yet effective approaches for constructing prediction bounds based on outcome regression: Outcome Regression (OR), and Calibrated Outcome Regression (COR). The first estimates the conditional quantile function of the primary event time at the desired nominal level, while the second calibrates the estimated conditional quantile function of the survival time to achieve the desired marginal coverage guaranty. Although the validity of this method is contingent on a consistent estimator of the survival function model, it provides a computationally efficient alternative to the IPCW and AIPCW approaches. To the best of our knowledge, this simple calibration-based technique has not been previously explored in the context of conformal predictive inference for censored time-to-event data.

\section{IPCW and AIPCW Methods}

\subsection{IPCW Method}
In order to ensure identifiability, throughout, we restrict the degree of dependence between the survival and censoring times, by invoking a standard  conditional independence censoring assumption  \citep{Kalb}. 
\begin{assumption}[Conditional independent censoring]\label{Assum:CIC} Event times and censoring times are independent conditional on the observed covariates, i.e.,
$T \perp \!\!\! \perp C \mid X.$
\end{assumption}
This assumption is analogous to missingness at random (MAR) in the missing data literature and to unconfoundness in causal inference. It is appropriate in practice only to the extent that baseline covariates reasonably capture the dependence between the primary outcome and the censoring mechanism.

We propose a novel approach to construct valid PAC type LPBs for the survival time of future observations, satisfying Definition \ref{def:PAC}. The key idea behind our method is to calibrate the coverage of prediction sets for the underlying time-to-event outcome in view, by incorporating IPCW to complete-cases to account for selection bias due to censoring, using for weight the reciprocal of the estimated conditional survival function of the censoring mechanism given covariates, evaluated at the observed failure time; therefore allowing us to properly estimate and thus calibrate the coverage probability. In fact, for a given function $R(x,t)$ possibly based on an independent sample, and for any $\beta$, under Assumption \ref{Assum:CIC},
\begin{equation}\label{eq:IPCW}
    \mathbb{P}\left(R(X,T)\ge \beta\right) = \mathbb{E}\left[\frac{\Delta \mathbf{1}\{R(X,T)\ge \beta\}}{S_{C\mid X}^\star(T\mid X)} \right].
\end{equation}
\cite{robins1992recovery} prove this result for a general full data estimating function. For completeness, the proof in our setting is given in 
Section~\ref{App:1} of the Supplementary Material. If $R(X,T)=\hat{L}(X)-T$, for some function $\hat{L}$ estimated from the observed data, then a proper choice of $\beta$ yields a LPB on $T$ of the form $\hat L(X)-\beta$. Similarly, if $R(X,T)=T-\hat{U}(X)$, for some function $\hat{U}$ estimated from the observed data, then we recover an upper predictive bound $\hat U(X)+\beta$. Moreover, two-sided predictitve bounds can be constructed using, for example, $R(X,T)=\max\{\hat{L}(X)-T, T-\hat{U}(X)\}$. In general, $R(X,T)$ can take any user-specified form, however, to simplify the exposition, consistent with prior works, we focus on LPBs.

Implementing the proposed IPCW estimator of the coverage probability first requires obtaining an estimator of the conditional survival curve for the censoring time $\hat{S}_{C\mid X}(\cdot\mid \cdot)$, and identifying a suitable, potentially covariate dependent, candidate for the LPB $\hat L$. A natural candidate is given by an estimate of the conditional quantile function $\hat{q}_{T\mid X}(\cdot\mid \cdot)$, which can be obtained from the fitted survival curve for $T\mid X$, $\hat{S}_{T\mid X}(\cdot\mid \cdot)$. Empirically, this translates into solving for the ``pseudo-quantile" level $\beta \in [0,1]$ -- the largest number to satisfy 
\[
\mathbb{E}\left[\frac{\Delta \mathbf{1}\{T\ge \hat{q}_{T\mid X}(\beta\mid X)\}}{\hat{S}_{C\mid X}(T\mid X)} \right] \ge 1-\alpha.
\]
We refer to $\beta$ as \emph{pseudo-quantile} level as the approach in fact does not rely on $\hat{q}_{T\mid X}(\cdot\mid \cdot)$ to be consistent for the true conditional quantile function in order for the resulting prediction interval to be  well-calibrated; rather, any working model, possibly incorrect, for the conditional quantile function is guaranteed to produce a prediction set which has correct marginal coverage. However, the optimal or oracle choice of $\hat{q}_{T\mid X}(\beta\mid X)$, assuming the survival curve of censoring defining the weights matches the truth, entails the true conditional quantile function, because then,  $\beta \equiv \alpha$ by definition of the conditional quantile, ensuring not only correct marginal coverage but also correct conditional coverage, a clear improvement in coverage control. This robustness property is a key contribution of our proposed approach.

For the implementation of this method, we consider the computationally efficient variant of conformal prediction, namely split conformal prediction, which relies on dividing the data into two independent splits. Specifically, we split the data $\mathcal{D}$ into a training set $\mathcal{D}_1$ and a calibration set $\mathcal{D}_2$, with respective sets of indices $\mathcal{I}_1$ and $\mathcal{I}_2$. Although our methods and results are formally stated for arbitrary sizes of $\mathcal{D}_1$ and $\mathcal{D}_2$, we assume henceforth that the two splits are of comparable sizes when discussing the rates of convergence of error terms. For example, by assuming that $\left|\mathcal{D}_1\right| / \left|\mathcal{D}_2\right| \in (0, \infty)$ as $n \to \infty$, it becomes feasible to express various error convergence rates in terms of $n$.

On the training set, we fit the conditional survival curves for both the survival time and the censoring time. We then estimate the coverage probability on the calibration set using the following IPCW estimator:
\begin{equation*}
\hat{P}_{\mathrm{HT-IPCW}}(\beta)=\frac{1}{\left| \mathcal{D}_2\right| }\sum_{i\in \mathcal{I}_2} \frac{\Delta_i\mathbf{1}\{T_i\ge \hat{q}_{T\mid X}(\beta\mid X_i)\}}{\hat{S}_{C\mid X}(T_i\mid X_i)},
\end{equation*}
for all $\beta\in[0,1]$. Finally, we select the optimal $\beta$ as
\[
\hat{\beta}_{\mathrm{HT-IPCW}} = \sup\left\{\beta\in[0,1]:\hat{P}_{\mathrm{HT-IPCW}}(\beta)\ge 1-\alpha\right\},
\]
and obtain the LPB as $\hat{L}(\cdot)=\hat{q}_{T\mid X}(\hat{\beta}_{\mathrm{HT-IPCW}}\mid \cdot)$. The classical IPCW estimator can be viewed as a form of Horvitz–Thompson estimator (HT-IPCW), a well-known inverse probability weighted estimator commonly used in the survey sampling literature, which computes the sample average normalizing by the sample size  \citep{Horvitz}. However, if estimated survival probabilities defining the weights $\hat{S}_{C\mid X}(T_i\mid X_i)$ are relatively small, the estimator may become unstable, leading to increased variability. To address this issue, we consider the Hájek version of IPCW estimator \citep{hajek1971comment, basu1971essay}, a well-known ratio-adjusted version of the Horvitz–Thompson estimator. The Hájek IPCW estimator normalizes by the sum of the inverse probability weights, such that the coverage probability is estimated as
\begin{equation*}
\hat{P}_{\mathrm{IPCW}}(\beta)=\frac{\sum_{i\in \mathcal{I}_2 } {\Delta_i\mathbf{1}\{T_i\ge\hat{q}_{T\mid X}(\beta\mid X_i)\}}/{\hat{S}_{C\mid X}(T_i\mid X_i)}}{\sum_{i\in \mathcal{I}_2 }{\Delta_i}/{\hat{S}_{C\mid X}(T_i\mid X_i)}},
\end{equation*}
and the optimal $\beta$ is given by
$\hat{\beta}_{\mathrm{IPCW}}= \sup\{\beta\in[0,1]:\hat{P}_{\mathrm{IPCW}}(\beta)\ge 1-\alpha\}$.
The Hájek estimator is approximately unbiased and often yields lower variance than the Horvitz-Thompson estimator \citep{datta}.
An equivalent estimating equation representation of the Hájek estimator is:
\[
\mathbb{E}\left[\frac{\Delta \left\{\mathbf{1}\{T\ge \hat{q}_{T\mid X}(\beta\mid X)\}-(1-\alpha)\right\}}{\hat{S}_{C\mid X}(T\mid X)} \right]\ge 0.
\]
Thus, we may estimate the optimal $\beta$ as
$\hat{\beta}_{\mathrm{IPCW}}= \sup\{\beta\in[0,1]:\hat{W}(\beta)\ge 0\}$,
where
\[
\hat{W}(\beta) = \frac{1}{\left| \mathcal{D}_2\right| } \sum_{i \in \mathcal{I}_2} \frac{\Delta_i \left\{\mathbf{1}\{T_i \ge \hat{q}_{T\mid X}(\beta\mid X_i)\} - (1-\alpha)\right\}}{\hat{S}_{C\mid X}(T_i \mid  X_i)}.
\]
Algorithm~\ref{algo:IPCW} in the Supplementary Material provides a detailed description of this procedure.

The method described above can be extended to a general non-conformity score computed on an independent sample. Let $R:\mathbb{R}^{d+1}\rightarrow\mathbb{R}$ be an arbitrary function defining a given non-conformity score. One may think of $R$ as a fixed non-stochastic function. This framework allows us to select $R$ to be of a simple form, such as $T$, or of more general form determined by the data, depending on the problem. If the LPB $\hat{L}$ is chosen to be the estimated conditional quantile of $T$ given $X$, then $R$ is defined through
\begin{equation}\label{eq:conditional-quantile-non-conformity-score}
\left\{R(X,T)\ge\beta\right\}= \left\{T\ge \hat q_{T\mid X}(\beta\mid X)\right\},
\end{equation}
implying that under correct specification of $\hat q_{T\mid X}$, $R(X,T)$ matches the estimated conditional cumulative distribution function $\hat F_{T\mid X}(T\mid X)$. In general, for any non-conformity score $R$, there is a corresponding LPB $\hat{L}$, which one may recover from an expression for $R$. Heretoafter, we will use $R$ to indicate that our methodology allows for an arbitrary choice of non-conformity score. %and does not depend on its choice. 
The detailed description of this procedure for a general non-conformity score $R$ is given in Algorithm~\ref{algo:IPCW-R} in the Supplementary Material.

In Theorem \ref{thm:valid-IPCW}, we establish the asymptotic marginal coverage guarantee for the LPB constructed through the procedure described above. We present the result in the case of a general non-conformity score $R$. In addition, when the non-conformity score is chosen as in~\eqref{eq:conditional-quantile-non-conformity-score}, we further obtain an asymptotic conditional coverage guarantee. The following positivity assumption is required on the conditional survival curve of $C\mid X$ and its estimator. 
\begin{assumption}[Positivity]\label{Assum:bound}
    There exists $0< s_0<\infty$ such that $\hat S_{C\mid X}(T\mid X)\wedge S_{C\mid X}^\star(T\mid X) > 1/s_0,$   
    a.s. under $P_{(X,T)}$.
\end{assumption}
This is a standard positivity condition on which much of the semiparametric efficiency theory for censored data is based on. It ensures that, for all the results achieved $T$ and $X$, there is a positive probability of observing an individual with $C \geq T$. For a detailed discussion, refer to Section 3.3 of \cite{laan2003unified}. Let $L^2$ denote the space of all square-integrable functions with respect to the probability measure $\mu$ of a random variable $Z$, where $\mu(B) = \mathbb{P}(Z \in B)$ for any Borel set $B$. We denote by $ \| f\| _{L^2} $ the $L^2$-norm of a function $f(\cdot)$, defined as
\[
\| f\| _{L^2} = \left( \int f^2(z)\mu(dz)\right)^{1/2}= \left(\mathbb{E}\left[f^2(Z)\right]\right)^{1/2}.
\]
Let $\| f\| _{\infty} = \sup_z |f(z)| $ denote the sup norm of a function $f(\cdot)$. 
\begin{theorem}\label{thm:valid-IPCW}
Let $\epsilon\in(0,1)$ be fixed. There exists a universal constant $K$ such that under Assumptions \ref{Assum:CIC} and \ref{Assum:bound}, with probability at least $1-\epsilon$ over $\mathcal{D}$
\begin{align}\label{eq:marg_cov}
    \mathbb{P}\left(R(X,T)\ge \hat{\beta}_{\mathrm{IPCW}} \mid \mathcal{D}\right) &\ge 1-\alpha-s_0^2\| \hat{S}_{C\mid X} - S_{C\mid X}^\star\|_{L^2} -s_0 \frac{\left(-\log\epsilon/2\right)^{1/2} + K}{\left| \mathcal{D}_2\right|^{1/2}},
\end{align}
where the probability $\mathbb{P}$ is taken with respect to a new data point $(X,T)\sim P_{(X,T)}$. Furthermore, if the non-conformity score is chosen to satisfy~\eqref{eq:conditional-quantile-non-conformity-score}, and if
the distribution of $T\mid X$ is continuous, then with probability at least $1-\epsilon$ over $\mathcal{D}$
\begin{align}\label{eq:cond_cov}
    \mathbb{P}\left(T\ge \hat{q}_{T\mid X}(\hat{\beta}_{\mathrm{IPCW}}\mid X) \mid X,\mathcal{D}\right) &\ge 1-\alpha-s_0^2\| \hat{S}_{C\mid X} - S_{C\mid X}^\star\|_{L^2} -2\|\hat{S}_{T\mid X} - S_{T\mid X}^\star\|_\infty\notag\\
    &\quad - s_0 \frac{\left(-\log\epsilon/2\right)^{1/2} + K}{\left| \mathcal{D}_2\right|^{1/2}},
\end{align}
where the probability $\mathbb{P}$ is taken with respect to a new data point $(X,T)\sim P_{(X,T)}$.
\end{theorem}
The proof of Theorem \ref{thm:valid-IPCW} is deferred to Section~\ref{App:2} of the Supplementary Material. As the size of the calibration set $\left| \mathcal{D}_2\right| $ goes to $\infty$, the last term on the right-hand side of~\eqref{eq:marg_cov} vanishes. The second error term converges to zero if $\hat{S}_{C\mid X}$ is consistent for $S_{C\mid X}^\star$. In other words, the prediction region for a new $X$, given by $\{t:R(X,t)\ge\hat{\beta}_{\mathrm{IPCW}}\}$, is asymptotically $(\alpha,\epsilon)$-PAC, implying that the corresponding $\hat{L}$ is an asymptotic $(\alpha,\epsilon)$-PAC LPB for $T$. Moreover, the prediction set enjoys asymptotic conditional coverage when the non-conformity score is defined as in~\eqref{eq:conditional-quantile-non-conformity-score}. In other words, asymptotic conditional validity is achieved provided that both conditional survival functions for the censoring mechanism and the time-to-event are consistently estimated.

\begin{remark}\label{rem:comp_ind}
In the special case of completely independent censoring, where the censoring time $C$ is independent of $(T,X)$, the estimation of the censoring mechanism simplifies significantly. In this scenario, the censoring survival function $S_{C}^\star(t)$ depends only on $t$ and can be consistently estimated at a root-$n$ rate using standard nonparametric methods, such as the Kaplan-Meier estimator \citep{Kaplan,fleming2013counting}. This assumption, while more restrictive, is commonly satisfied in randomized trials, at least within treatment arms, where the censoring mechanism may reasonably be assumed to be independent of other variables by careful study design \citep{pocock2013clinical, collett2023modelling}. Consequently, the IPCW framework becomes even more robust and efficient in such settings, further strengthening its practical applicability.
\end{remark}

In short, the procedure outlined in this section constructs an asymptotically $(\alpha,\epsilon)$-PAC LPB $\hat{L}(\cdot)$ for the right-censored survival time $T$. Moreover, two-sided bounds can also be obtained if the non-conformity score is chosen appropriately: $R(X,T) = |2\widehat{S}_{T|X}(T|X) - 1|$~\citep{chernozhukov2021distributional}.

\subsection{AIPCW method}
The IPCW-based approach described in the previous section relies on weighting observations with observed failure time (i.e., the complete-cases) by the inverse of the probability of remaining uncensored up to the time they experience the primary outcome, thereby correcting for any potential bias due to censoring. While this method can be effective under fairly reasonable conditions, it can be inefficient by virtue of not making efficient use of information contained in censored observations %either the survival model or the censoring model is misspecified 
and may not produce well-calibrated prediction sets if either the estimator of the censoring mechanism converges at slow rates, or fails to be consistent. To address these limitations, we extend our framework by introducing an augmentation to IPCW that recovers information from censored observations by leveraging a model for the outcome of interest. 

The AIPCW method has attractive theoretical appeal over the IPCW approach, by producing a doubly robust estimator, which means that the estimator remains consistent if either the survival model or the censoring model is correctly specified (but not necessarily both). This is an immediate consequence of its mixed bias error as shown in the result below. This robustness can be particularly valuable in survival analysis, even if machine learning or nonparametric methods are used to estimate unknown regression functions, as the AIPCW error rate is then guaranteed to be of smaller order of magnitude than that of IPCW.

The efficiency and robustness properties of AIPCW are rooted in modern semiparametric efficiency theory, where influence functions are central to constructing efficient estimators. Briefly, according to a result due to \cite{robins1992recovery}, the efficient influence function of a regular parameter defined on a nonparametric model subject to censoring at random can be obtained as the projection of the IPCW of the full data efficient influence function onto the ortho-complement to the tangent space for the censoring mechanism assumed to satisfy Assumption~\ref{Assum:bound} and otherwise unrestricted. The tangent space for a semiparametric model is technically defined as the closed linear span of scores of all regular parametric sub-models contained in the model. \cite{robins1992recovery} proved that under Assumptions \ref{Assum:CIC} and \ref{Assum:bound}, the tangent space for the censoring mechanism is given by
\[
\mathcal{T}=\mbox{closure}\left\{{\int h\left(u,X\right) dM_{C\mid X}^\star\left(u\mid X\right):h\in L^2}\right\},
\]
where
$dM_{C\mid X}^\star( u\mid X) =dN_C(u) - \mathbf{1}\{Y\geq u\} d\Lambda^\star_{C\mid X}( u\mid X)$
is the martingale difference process with $dN_C(u)=\mathbf{1}\{ Y\in du,\Delta =0\}$ being the increment of the censoring counting process, and $\Lambda^\star_{C\mid X}\left( u\mid X\right) = -\log(S_{C\mid X}^\star\left( u\mid X\right))$ being the true censoring cumulative hazard function. Let $W(\cdot)$ be the IPCW estimating function: 
\[
W(\beta)=\frac{\Delta\left\{\mathbf{1}\{R(X,T)\ge \beta\}-(1-\alpha)\right\}}{S_{C\mid X}^\star\left(T\mid X\right)}.
\]
Then, as shown in \citet{robins1992recovery}, the orthogonal projection of $W$ onto the censoring tangent space is given by
\[
\Pi\left(W(\beta)\mid \mathcal{T}\right)=-\int \frac{\mathbb{E}\left[ \mathbf{1}\{ R(X,T)\ge \beta \} \mid X,T\geq u\right] - (1-\alpha) }{S_{C\mid X}^\star\left(u\mid X\right)} dM_{C\mid X}^\star\left( u\right). 
\]
A self-contained proof of this result is provided in Section \ref{App:aipcw-projection} of the Supplementary Material. The resulting AIPCW moment equation is thus given by the residual of the projection of the IPCW estimating function onto the censoring tangent space. As a result, we propose to identify the optimal quantile value $\beta\in[0,1]$ as the largest number that satisfies
\begin{equation}\label{eq:estIF}
\mathbb{E}\left[\frac{\Delta \left\{\mathbf{1}\{R(X,T)\ge \beta\}-(1-\alpha)\right\}}{\hat{S}_{C\mid X}(T\mid X)}+\int \frac{\hat\eta(\beta,u\mid X) - (1-\alpha) }{\hat S_{C\mid X}\left(u\mid X\right)} d\hat M_{C\mid X}\left( u\mid X\right) \right]\ge0,
\end{equation}
where $\hat\eta(\beta,u\mid X)$ is an estimator of $\eta^\star(\beta,u\mid X)=\mathbb{E}\left[ \mathbf{1}\{R(X,T)\ge \beta\} \mid X,T\geq u\right]$. There are two aspects with the AIPCW moment equation that require further elaboration for practical implementation: (i) estimation of the additional nuisance component $\hat{\eta}$; and (ii) computation of the integral with respect to the martingale difference process. Similarly to the IPCW method, the data $\mathcal{D}$ is divided into a training set $\mathcal{D}_1$ and a calibration set $\mathcal{D}_2$, with respective index sets $\mathcal{I}_1$ and $\mathcal{I}_2$. We use $\mathcal{D}_1$ to estimate all nuisance functions.

For (i), note that when, as we suppose in the following exposition, the estimated conditional quantile of $T\mid X$ (from an independent sample) is used to define the non-conformity score (as in~\eqref{eq:conditional-quantile-non-conformity-score}),
\begin{align}\label{eq:conditional-quantile-eta}
    \eta^\star(\beta,u\mid X) &= \mathbb{E}\left[ \mathbf{1}\{T\ge \hat{q}_{T\mid X}(\beta\mid X)\} \mid X,T\geq u, \mathcal{D}_1\right]\notag\\
    &=\mathbb{P}\left(T\ge \hat{q}_{T\mid X}(\beta\mid X)\mid X,T\geq u, \mathcal{D}_1\right)\notag\\
    &=\frac{\mathbb{P}\left( T\ge \hat{q}_{T\mid X}(\beta\mid X), T\geq u \mid X, \mathcal{D}_1 \right)}{\mathbb{P}\left( T\ge u \mid X, \mathcal{D}_1\right)}\notag\\
    &=\frac{S_{T\mid X}^\star\left( \max\{\hat{q}_{T\mid X}(\beta\mid X), u\} \mid X \right)}{S_{T\mid X}^\star\left( u \mid X\right)}.
\end{align}
Thus, in this case, an estimator of $\eta^\star(\beta,u\mid X)$ is conveniently expressed as:
\[
\hat\eta(\beta,u\mid X) = \frac{\hat S_{T\mid X}\left( \max\{\hat q_{T\mid X}(\beta\mid X), u\} \mid X \right)}{\hat S_{T\mid X}\left( u \mid X\right)}.
\]

For (ii), note that when evaluated in the observed sample, the integral in equation~\eqref{eq:estIF} often reduces to a finite sum. In fact, for several estimators, like random survival forests, $\hat{S}_{C\mid X}$ is piecewise constant with jumps at observed censoring times, implying that $\hat{\Lambda}_{C\mid X}$ is also piecewise constant and the martingale difference process is a discrete measure. In such cases, denoting the observed ordered censoring times as $u_{(1)},\dots,u_{(Q)}$ for some finite $Q$, and setting $u_{(0)}=0$, we have:
\begin{align*}
&\int\frac{\hat\eta(\beta,u\mid X)-(1-\alpha)}{\hat{S}_{C\mid X}(u\mid X)}d\hat M_{C\mid X}\left(u\mid X\right) \\
&= \sum_{k=1}^Q \frac{\hat\eta(\beta,u_{(k)}\mid X)-(1-\alpha)}{\hat{S}_{C\mid X}(u_{(k)}\mid X)}\left\{\hat M_{C\mid X}\left(u_{(k)}\mid X\right)-\hat M_{C\mid X}\left(u_{(k-1)}\mid X\right)\right\},
\end{align*}
where $\hat M_{C\mid X}\left(u\mid X\right)=\mathbf{1}\{ Y\le u,\Delta =0\} - \hat\Lambda_{C\mid X}\left(Y\land u\mid X\right)$. 
Note that, more generally, for estimators of the censoring mechanism that rely on smoothing the baseline hazard, such as kernel smoothing, the integral may not naturally reduce to a finite sum over censoring times; however, the latter may still be approximated by an alternative (weighted) finite sum via, say, Gaussian quadrature or analogous numerical integration techniques.

On the training set $\mathcal{D}_1$, the conditional survival curves for both the survival time and the censoring time are estimated. Subsequently, on the calibration set $\mathcal{D}_2$, we compute the IPCW term as
\begin{equation}
\hat{W}(\beta)=\frac{1}{\left| \mathcal{D}_2\right| }\sum_{i\in \mathcal{I}_2} \frac{\Delta_i\left\{\mathbf{1}\{T_i\ge \hat{q}_{T\mid X}(\beta\mid X_i)\}-(1-\alpha)\right\}}{\hat{S}_{C\mid X}(T_i\mid X_i)},
\end{equation}
and the augmentation term as
\begin{align*}
    \hat{\Pi}(\beta)&=\frac{1}{\left| \mathcal{D}_2\right| }\sum_{i\in \left| \mathcal{D}_2\right| } \sum_{k=1}^Q \frac{\hat\eta(\beta,u_{(k)}\mid X_i)-(1-\alpha)}{\hat{S}_{C\mid X}(u_{(k)}\mid X_i)} \left\{\hat M_{C\mid X}(u_{(k)}\mid X_i)-\hat M_{C\mid X}(u_{(k-1)}\mid X_i)\right\},
\end{align*}
for all $\beta\in[0,1]$. The optimal $\beta$ is then estimated as
$\hat{\beta}_{\mathrm{AIPCW}} = \sup\{\beta\in[0,1]:\hat{W}(\beta) + \hat{\Pi}(\beta)\ge 0\}$.
The detailed steps of this procedure are provided in Algorithm~\ref{algo:AIPCW} in the Supplementary Material.
%Algorithm~\ref{algo:AIPCW} outlines the detailed steps of this procedure. 
As with the IPCW-based method, this procedure can be generalized to use any arbitrary non-conformity score $R(X,T)$, with the corresponding steps given in Algorithm~\ref{algo:AIPCW-R} in the Supplementary Material. 

The rest of this section establishes the asymptotic validity and double robustness of our proposed AIPCW-based approach. The efficient influence function can readily be derived in analogy to the derivation above, as the projection of an IPCW moment function onto the ortho-complement of the censoring tangent space: 
\begin{align}\label{eq:IF}
    \mathrm{IF}_\beta(S_{C\mid X},\eta)&=\frac{\Delta}{S_{C\mid X}(T\mid X)} \left\{\mathbf{1}\{R(X,T)\ge\beta\}-(1-\alpha)\right\}\notag \\
    &\quad+\int \left\{\eta(\beta,u\mid X) - (1-\alpha)\right\} \frac{dM_{C\mid X}(u\mid X)}{S_{C\mid X}(u\mid X)}.
\end{align}
For clarity, the explicit dependence on $X, T$ and $R$ is suppressed. Additionally, note that the dependence on $dM_{C\mid X}$ is captured through the dependence on $S_{C\mid X}$. In Lemma \ref{lem:IF}, we provide an alternative expression for the efficient influence function of $\beta$ which is instrumental in proving the key double robustness result outlined in Theorem \ref{thm:dbl_rob}.
\begin{lemma}\label{lem:IF} 
Under Assumption \ref{Assum:CIC}, 
\begin{align*}
    \mathrm{IF}_\beta(S_{C\mid X},\eta)&=\mathbf{1}\{ R(X,T)\ge\beta\}-(1-\alpha) \\
    &\quad-\int \left\{\mathbf{1}\{R(X,T)\ge\beta\}  -\eta(\beta,u\mid X) \right\}\frac{dM_{C\mid X}(u\mid X)}{S_{C\mid X}(u\mid X)}.
\end{align*}
\end{lemma}
This expression holds for any choice of functions $\eta$ and $S_{C\mid X}$, whether estimated or evaluated at their true value. The proof of Lemma \ref{lem:IF} can be found in Section~\ref{App:3} of the Supplementary Material.

The main double robustness property of the derived efficient influence function of $\beta$ is established below. Before presenting the result, we introduce some additional notation. For any function $f^\star:\mathbb{R}^{d+1}\to \mathbb{R}$ with estimate $\hat{f}$, define
\begin{gather*}
\mathrm{Err}_{1,f} := \|\hat{f}(Y\mid X) - f^\star(Y\mid X)\|_{L^2}, \quad
\mathrm{Err}_{2,f}(u) := \|\hat{f}(u\mid X) - f^\star(u\mid X)\|_{L^2},\\[0.8em]
\mathrm{Err}_{3,f}(u) := \left\|\frac{d\hat{f}(u\mid X)}{du} - \frac{df^\star(u\mid X)}{du}\right\|_{L^2}.
\end{gather*}
Let $\mathrm{Err}_{j,{\eta}^\star_{\beta}}(\cdot), 1\le j\le 3$ denote the errors in estimating $\eta^\star(\beta,\cdot\mid\cdot)$, and similarly, let $\mathrm{Err}_{j,\Lambda_{C\mid X}^\star}(\cdot), 1\le j\le 3$ denote the errors in estimating $\Lambda_{C\mid X}^\star(\cdot\mid\cdot)$.
\begin{theorem}\label{thm:dbl_rob} Under Assumptions \ref{Assum:CIC} and \ref{Assum:bound}, for any estimated (from $\mathcal{D}_1$) functions $\hat S_{C\mid X}$ and $\hat\eta$, and for any $\beta\in[0,1]$, the following inequality holds
\begin{align}\label{eq:DR}
    &\left|\mathbb{E}\left[\mathrm{IF}_\beta(\hat S_{C\mid X}, \hat\eta)\big|\mathcal{D}_1\right]-\mathbb{E}\left[\mathrm{IF}_\beta(S_{C\mid X}^\star,\eta^\star)\right]\right|\notag<s_0 \min\bigg\{\int \mathrm{Err}_{2,{\eta}^\star_{\beta}}(u) \mathrm{Err}_{3,\Lambda_{C\mid X}^\star}(u)du,\notag\\
    &\quad\mathrm{Err}_{1,{\eta}^\star_{\beta}} \mathrm{Err}_{1,\Lambda_{C\mid X}^\star} + \int \mathrm{Err}_{3,{\eta}^\star_{\beta}}(u) \mathrm{Err}_{2,\Lambda_{C\mid X}^\star}(u)du\bigg\}.
\end{align}
\end{theorem}
This result implies that the influence function $\mathrm{IF}_\beta(\hat S_{C\mid X}, \hat\eta)$ is asymptotically unbiased as long as either $S_{C\mid X}^\star$ or $\eta^\star$ is estimated consistently. The proof of Theorem \ref{thm:dbl_rob} is detailed in Section~\ref{App:4} of the Supplementary Material. 

\begin{remark}
    As mentioned in Remark \ref{rem:comp_ind}, under completely independent censoring, the censoring survival curve $S_C^\star(u)$, and therefore the cumulative hazard function $\Lambda_C^\star(u)$, can be consistently estimated at a root-$n$ rate. Hence, $\|  \hat{\Lambda }_{C}(u) -\Lambda _{C}(u) \| _{L^2}=O_{p}\left( n^{-1/2}\right)$, which implies that the error bound of the AIPCW estimator is negligible and is guaranteed to be better than that of the IPCW estimator.
\end{remark}

\begin{remark}
    When the estimated conditional quantile of $T\mid X$ is used to define the non-conformity score (as in~\eqref{eq:conditional-quantile-non-conformity-score}), $\eta^\star$ takes the specific form described in~\eqref{eq:conditional-quantile-eta}, which involves the ratio of $S_{T\mid X}^\star$ evaluated at two different time points. Therefore, under a positivity assumption for the time-to-event, analogous to that of the censoring mechanism in Assumption \ref{Assum:bound}, the error in estimating $\eta^\star$ can be bounded by the corresponding error in estimating $S_{T\mid X}^\star$, scaled by the positivity constant.
    %the assumption that $S_{T\mid X}^\star$ and $\hat{S}_{T\mid X}$ satisfy a positivity condition analogous to that of the censoring mechanism in Assumption \ref{Assum:bound}, 
\end{remark}

\begin{remark}
    The coverage error bound for the AIPCW estimator depends not only on the estimation errors of the nuisance functions themselves but also on the estimation accuracy of their derivatives. In general, the convergence rate of a derivative of a function is slower than that of the function itself, depending on the smoothness of the function. Observe that $\eta^\star$ (for most scores) can be written in terms of $S^\star_{T|X}$, as shown in~\eqref{eq:conditional-quantile-eta}. Also, $\Lambda^\star_{C|X}$ is a smooth function of $S^\star_{C|X}$. Hence, the typical rates of convergences of $\mathrm{Err}_{j,\eta^\star}$ and $\mathrm{Err}_{j,\Lambda^\star}$ are the same for each $j\in\{1, 2, 3\}$. As with any non-parametric function estimation problem, the actual rates of convergences depend on the smoothness assumptions on these functions. Under (semi-)parametric models, faster rates are possible. For instance, in the Cox proportional hazards model, with the cumulative baseline hazard estimated via the Breslow estimator, the cumulative hazard converges at the standard parametric rate $O_p(n^{-1/2})$ for a fixed time point, under standard regularity conditions \citep{tsiatis1981large}. 
    Its derivative, the hazard function, does not have a well-defined convergence rate since the Breslow estimator is a step function that only increases at observed event times. To address the non-smoothness of the Breslow estimator, smooth estimators are typically used. They can have different convergence rates depending on the specific smoothing technique and the smoothness of the underlying true function. Under twice differentiability, the rate is $O_p(n^{-2/(d+4)})$ \citep{lopuhaa2017smooth}.
\end{remark}

The asymptotic coverage guarantee for the LPB constructed using the AIPCW-based approach, for a general non-conformity score $R$, is given in the following theorem, which represents a key novelty of the paper as, to the best of our knowledge, the first marginal coverage guarantee for predictive inference for a right-censored time-to-event outcome, with mixed bias of coverage error rate.
\begin{theorem}\label{thm:valid-AIPCW}
Let $\epsilon\in(0,1)$ be fixed. There exists a universal constant $K$ such that under assumptions \ref{Assum:CIC} and \ref{Assum:bound}, with probability at least $1-\epsilon$ over $\mathcal{D}$
\begin{align*}
    \mathbb{P}\left(R(X,T)\ge\hat{\beta}_{\mathrm{AIPCW}}\mid \mathcal{D}\right)&> 1-\alpha-\left(s_0+2\max\left\{1,s_0-1\right\}\right)\frac{\left(-\log\epsilon/2\right)^{1/2} + K}{\left| \mathcal{D}_2\right|^{1/2}}\\
    &\quad- s_0 \sup_{\beta\in[0,1]}\min\bigg\{\int \mathrm{Err}_{2,{\eta}^\star_{\beta}}(u) \mathrm{Err}_{3,\Lambda_{C\mid X}^\star}(u)du,\\
    &\quad\mathrm{Err}_{1,{\eta}^\star_{\beta}} \mathrm{Err}_{1,\Lambda_{C\mid X}^\star} 
    + \int \mathrm{Err}_{3,{\eta}^\star_{\beta}}(u) \mathrm{Err}_{2,\Lambda_{C\mid X}^\star}(u)du\bigg\},
\end{align*}
where the probability $\mathbb{P}$ is taken with respect to a new data point $(X,T)\sim P_{(X,T)}$.
\end{theorem}
The proof of Theorem \ref{thm:valid-AIPCW} is presented in Section~\ref{App:5} of the Supplementary Material.

\begin{remark}
    While the coverage guarantees for IPCW (Theorem \ref{thm:valid-IPCW}) and for AIPCW (Theorem \ref{thm:valid-AIPCW}) appear to involve different types of estimation errors for the censoring mechanism, with the former expressed in terms of the survival curve $S_{C\mid X}^\star$ and the latter in terms of the cumulative hazard function $\Lambda_{C\mid X}^\star$, these errors are closely related due to their smooth relationship, i.e., $\Lambda^\star_{C\mid X}\left( u\mid X\right) = -\log(S_{C\mid X}^\star\left( u\mid X\right))$. This implies that we can bound the error in estimating $S_{C\mid X}^\star$ with the error in estimating $\Lambda_{C\mid X}^\star$, and vice versa, ensuring that the rates of convergence for these estimation errors are of the same order.
\end{remark}

\subsection{Calibrated Outcome Regression Method}
As discussed in Section \ref{sec:intro_probsetup}, the oracle LPB for the survival time is given by the conditional quantile of $T \mid X$ at level $\alpha$, $q^\star_{T\mid X} (\alpha \mid \cdot)$. A naive approach to estimate an LPB is Outcome Regression (OR), which consists of directly estimating the conditional quantile $\hat q_{T\mid X}(\alpha \mid X)$. However, such an approach does not necessarily guarantee the desired marginal coverage probability. We introduce the Calibrated Outcome Regression (COR) method, which adjusts the quantile level to ensure proper calibration.
%Specifically, COR finds the largest level $\beta \in [0,1]$ such that:
%\[
%\mathbb{E}[\hat S_{T\mid X}(\hat q_{T\mid X}(\beta\mid X)\mid X)]\ge 1-\alpha.
%\]
This approach ensures that the estimated LPB satisfies the desired probabilistic coverage constraint by leveraging the estimated survival function of $T \mid X$. The implementation is straightforward and similar to that of IPCW and AIPCW methods. First, on the training set $\mathcal{D}_1$, we estimate the survival function $\hat{S}_{T\mid X}(\cdot \mid X)$ and use it to compute the empirical conditional quantiles $ \hat{q}_{T\mid X}(\beta \mid X)$ over a grid of values for $\beta$. Second, on the calibration set $\mathcal{D}_2$, we determine the optimal quantile level by solving:
\[
\hat{\beta}_{\mathrm{COR}} = \sup\left\{\beta\in[0,1]:\frac{1}{\left| \mathcal{D}_2\right| } \sum_{i \in \mathcal{I}_2} \hat S_{T\mid X}(\hat q_{T\mid X}(\beta\mid X_i)\mid X_i) \ge 1-\alpha\right\}.
\]
Finally, we output the LPB $\hat{L}(\cdot) = \hat{q}_{T\mid X}(\hat{\beta}_{\mathrm{COR}} \mid \cdot)$.

The OR and COR methods provide a computationally efficient way to estimate an LPB while ensuring calibrated marginal and conditional coverage. However, their validity relies on the assumption that $\hat{S}_{T\mid X}(\cdot \mid X)$ is a consistent estimator of the true survival function. This approach is similar in spirit to maximum likelihood estimation (MLE) for survival analysis, as it relies directly on modeling the conditional survival function. If the model is misspecified, the resulting predictions may be biased. The OR and COR methods account for dependence between censoring and the target outcome by conditioning on $X$ without directly modeling the censoring distribution. In contrast, IPCW and AIPCW methods explicitly account for censoring through inverse probability weighting. 
%To the best of our knowledge, this simple calibration-based technique has not been previously explored in the context of conformal predictive inference for censored time-to-event data. 
Although COR has a first-order bias, AIPCW benefits from a second-order mixed bias, making it theoretically more robust. In summary, COR offers a straightforward way to obtain well-calibrated prediction bounds using outcome regression, whereas AIPCW provides additional robustness by leveraging information from both the censoring and survival models. Importantly, we highlight that both OR and COR appear to be novel methods to the best of our knowledge.

\section{Simulation studies}\label{sec:simu}
\subsection{Simulation design and benchmark comparison methods}\label{sec:simu_design}
We conduct a comprehensive series of simulation studies to evaluate the performance of our proposed methods. The code for reproducing the results of our simulations 
can be found at \url{https://github.com/rebyfa98/DR_conformal_censored}.
For each simulation scenario, we generate 100 independent and identically distributed datasets. Each dataset is divided into three sets: the training set $\mathcal{D}_1$, the calibration set $\mathcal{D}_2$, and the test set $\mathcal{D}_3$, with $\left| \mathcal{D}_1\right| =\left| \mathcal{D}_2\right| =\left| \mathcal{D}_3\right| =1000$. 
We generate the full data where the event time $T$ is observed for all individuals in the test set $\mathcal{D}_3$. Although the full data is not utilized to implement our methods, it allows us to accurately compute the empirical coverage rate in the test set $\mathcal{D}_3$ as $
\left|\mathcal{D}_3\right|^{-1}\sum_{i\in\mathcal{I}_3} \mathbf{1}\{T_i\ge\hat{L}(X_i)\}$. Moreover, we compute the average estimated LPB in the test set $\mathcal{D}_3$ as
$\left| \mathcal{D}_3\right|^{-1} \sum_{i\in\mathcal{I}_3} \hat{L}(X_i)$.
In all experiments, we set the target coverage level to $1-\alpha = 0.9$. The grid of points ${\beta_j}$ is defined as equally spaced points $\{0.001j, j=0,\dots,1000\}$.

Both proposed methods are implemented three times, using three different estimators for the conditional survival curves: Cox proportional hazards regression model and Random Survival Forests (RSF), Super Learner (SL).
The Cox model, widely used in survival analysis, serves as a benchmark due to its popularity, with the cumulative baseline hazard function estimated using the Breslow estimator \citep{breslow1972contribution}. However, our primary focus is on nonparametric approaches, which do not impose restrictive modeling assumptions such as proportional hazards. RSF is a suitable choice for this purpose, leveraging its flexibility in modeling survival data. For further flexibility, we also employ SL, an ensemble method that optimally combines multiple survival algorithms by minimizing the cross-validated risk, thereby reducing the risk of overfitting in the final model \cite{Golmakani2020SuperLF}. To implement these estimators, we use available \texttt{R} packages: \texttt{survival} for the Cox model, \texttt{randomForestSRC} \citep{ishwaran2021fast} for RSF and \texttt{survSuperLearner} for SL. This SL algorithm optimally combines candidate models to estimate both the conditional survival and censoring functions simultaneously. The set of candidate learners includes Kaplan-Meier, Cox proportional hazards, Exponential regression, Weibull regression, Log-logistic regression, and Random Survival Forests. 
%No screening algorithm used in SL, i.e., we use all variables.
For the (COR) method and its non-calibrated version (OR), we directly apply the SL model to ensure greater flexibility.

We compare our methods against two groups of benchmarks. First, we consider ``naive'' quantile-based baselines that ignore censoring: quantile regression forests fit to the observed time $Y$ (QR-Y) and its conformalized version (CQR-Y), as well as quantile regression forests fit to the event time $T$ using only uncensored observations (QR-T), and its conformalized version (CQR-T). These approaches serve as reference points because they do not technically account for the censoring mechanism, despite relying on the same assumptions as our IPCW and AIPCW methods.  
Second, we compare against recent conformal survival methods, namely the approaches proposed by \citet{sesia2025doubly, qin2025conformal, si2025training, yi2025survival, meixide}. A brief description of each competing method, together with the implementation details used in our experiments, is provided in Section \ref{App:comp} of the Supplementary Material.

\subsection{Synthetic data}
We examine six synthetic experiments, outlined in Table~\ref{tab:simu_setting} of the Supplementary Material, designed to capture a range of censoring regimes and model complexities. 
In Settings 1--2, both the event time $T$ and censoring time $C$ follow covariate-dependent exponential models, implying that the Cox proportional hazards assumption holds. In Setting 1, we have completely independent censoring, i.e., $C$ is independent of $(T,X)$, and the observed censoring rate is about 30\%. In Setting 2, both $T$ and $C$ depend on nonlinear transformations and interactions of the covariates, yielding informative censoring, and the observed censoring rate is about 50\%. In Settings 3--6 we move beyond proportional hazards by using log-normal models, which allow us to assess the performance of nonparametric models such as RSF. In these settings, censoring is covariate-dependent, with observed censoring rates of approximately 62\% in Setting 3, 70\% in Setting 4, 66\% in Setting 5 and 50\% in Setting 6. Furthermore, Setting 5 includes nonlinear and interaction effects in both the event and the censoring mechanisms. Finally, Setting 6 introduces heteroskedasticity in the time-to-event model.

\subsection{Results}
\begin{figure}[t]
    \centering
    \includegraphics[width=\textwidth]{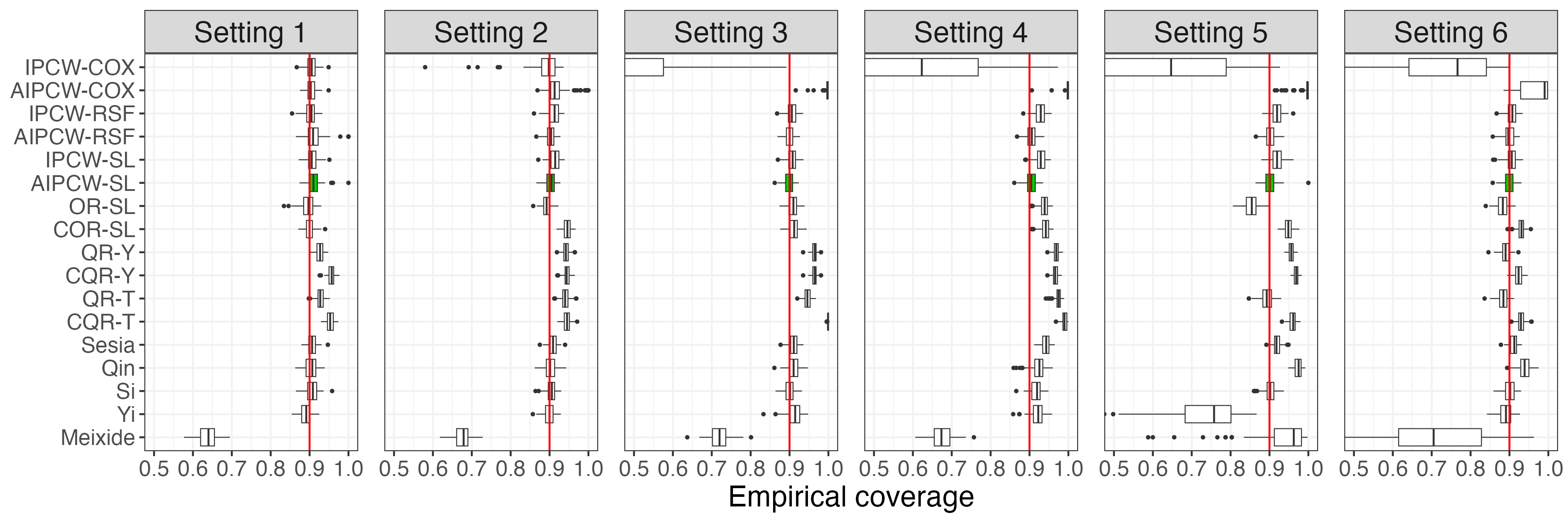}
    \caption{Distribution of empirical marginal coverage rates across 100 simulated test datasets for each method under Settings 1--6 (Table~\ref{tab:simu_setting}). The red vertical line denotes the 90\% target coverage. The recommended method AIPCW-SL is highlighted in green.
    }
    \label{fig:coverage_boxplot}
\end{figure}

\begin{figure}[t]
    \centering
    \includegraphics[width=\textwidth]{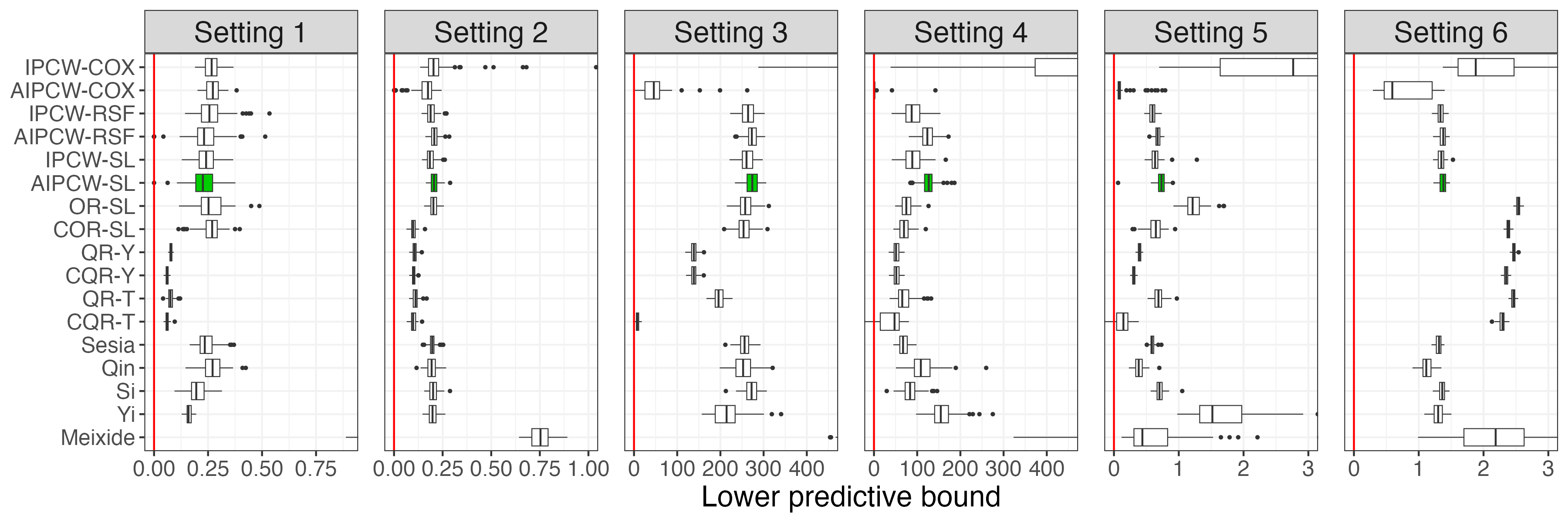}
    \caption{Distribution of the average estimated LPB in the test set across 100 simulated datasets for each method under Settings 1--6 (Table~\ref{tab:simu_setting}). The red vertical line at 0 marks the lower limit of the support of the time-to-event outcome. The recommended method AIPCW-SL is highlighted in green.}
    \label{fig:lpb_boxplot}
\end{figure}

Figure \ref{fig:coverage_boxplot} shows the boxplots illustrating the empirical coverage rates obtained from the 100 datasets in each setting, providing a visual comparison of the performance of all methods. For a closer examination of each simulation study, refer to the zoomed-in plots in Figures~\ref{fig:coverage_boxplot_setting1to3} and~\ref{fig:coverage_boxplot_setting4to6} in the Supplementary Material. It is important to note that our theoretical results are asymptotic, whereas these experiments are conducted on finite samples, thus some deviation from the target coverage is anticipated. In addition, Figure~\ref{fig:lpb_boxplot} displays the boxplots of the average estimated LPB values across the 100 datasets in each setting. In all plots, the method we recommend, AIPCW-SL, is highlighted in green.

Settings 1--2 satisfy proportional hazards, so the Cox-based nuisance models are appropriate. Indeed, in Setting 1 the IPCW/AIPCW methods attain coverage close to the target level, with similar LPBs. The outcome-regression approaches (OR/COR) also perform well because the data-generating model is simple and can be captured by a flexible learner such as SL. 
Setting 2 is more challenging since censoring is informative and both $T\mid X$ and $C\mid X$ depend on nonlinear transformations or interactions. Here, the IPCW/AIPCW methods remain well calibrated, whereas OR/COR lead to under- and over-coverage, respectively, reflecting sensitivity of outcome-regression approaches to errors in estimating the survival curve in this nonlinear informative-censoring regime.

In the misspecified Settings 3--6, the Cox-based IPCW and AIPCW procedures break down, producing substantial under- and over-coverage, respectively. When paired with flexible nuisance learners (RSF or SL), both IPCW and AIPCW recover near-nominal coverage. Moreover, AIPCW typically achieves coverage closer to the target and larger average LPBs than IPCW, which aligns with our theoretical results: the augmented construction is doubly robust and its coverage error is governed by a second-order mixed-bias remainder.
Outcome-regression methods perform well in Setting 3 but not in Settings 4--6, again due to the dependence on accurate estimation of the conditional survival curve, which becomes harder in high-dimensional, nonlinear or heteroskedastic regimes.

The naive methods (QR-Y, CQR-Y, QR-T, CQR-T) produce overly conservative bounds across all settings, as evidenced by high coverage rates and low estimated LPBs. Among recent competing methods, \citet{sesia2025doubly} and \citet{qin2025conformal} attain near-nominal coverage in Settings 1--3. However, their performance becomes worse in the more complicated Settings 4--6, where the structure and heavier censoring make it harder (i) to reliably perform imputation-based approaches \citep{sesia2025doubly}, and (ii) to stably bootstrap-calibrate prediction under limited tail information \citep{qin2025conformal}, leading to more conservative bounds. The concurrent EIF-based approach of \cite{si2025training} provides a valid alternative and behaves similarly to ours in most settings, but its more conservative APAC-style calibration can yield over-coverage in more difficult regimes, as in Setting 4, or slightly less informative LPBs than our AIPCW-SL method, as in Settings 5--6. Likewise, \cite{yi2025survival} performs well when the IPCW-quantile-regression combination can adequately capture the target quantile, but it becomes less reliable under stronger censoring model complexity, as in Setting 5, where it shows noticeable under-coverage. 
Finally, the method proposed by \citet{meixide}, which is designed for interval-censoring, is not directly competitive here.

Overall, our results indicate that the proposed AIPCW approach offers a favorable and comparatively stable trade-off between validity and bound sharpness across the range of scenarios considered. As expected, naive methods that ignore censoring tend to be conservative, producing high coverage and smaller LPBs. Competing methods can match our methods' performance in particular scenarios; however, in our experiments none of them simultaneously improves upon AIPCW-SL in calibration while also yielding more informative bounds across all settings. These findings support recommending AIPCW-SL as a practical default, as it maintains near-nominal coverage and informative LPBs.

Additional simulation results, including studies across different target coverage levels and training sample sizes, conditional coverage experiments, and further results on the performance of competing methods, are provided in Section~\ref{App:plots} of the Supplementary Material.

\section{Real Data Application}
\subsection{Data description}
We use data from a cohort of 1,240 patients with rheumatoid arthritis from the Wichita Arthritis Center, an outpatient rheumatology facility \citep{choi2002methotrexate}. The original study aimed to evaluate the impact of methotrexate, the most commonly used disease-modifying antirheumatic drug (DMARD), on patient mortality. Our goal is to apply our methods to estimate predictive bounds for the survival time to death.

During the data collection, each patient underwent longitudinal follow-up, with multiple assessments recorded over time. For our analysis, we focus on baseline covariates to account for patient characteristics at the start of the study. Among the 1,240 patients, 191 died during follow-up, resulting in an observed event rate of 15.4\%. The average follow-up time was 72.3 months. The covariates considered in our analysis include age, sex, disease duration, education level, source of enrollment, prednisone use, DMARD use, treatment contraindications, rheumatoid factor positivity, smoking status, and several clinical measures such as functional disability score, global assessment of disease score, and tender joint count. This dataset provides a robust framework for evaluating survival prediction methods in a real-world clinical setting where censoring is common.

\subsection{Performance assessment}
To evaluate the performance of our methods on this dataset, we estimate the empirical coverage of the predictive lower bounds using two different approaches to account for censoring: one based on inverse probability of censoring weighting (IPCW and AIPCW) and the other based on outcome regression. Unlike in the simulation study, where all event times were available for validation (for both uncensored and censored observations), here we only have access to event times for uncensored outcomes, thus requiring different techniques in estimating coverage.

The IPCW/AIPCW-based empirical coverage is computed by reweighting the observed failures to correct for censoring. Specifically, given an estimated optimal level $\beta$ and a test dataset $\mathcal{D}_3$ with index set $\mathcal{I}_3$, the IPCW-based empirical coverage is computed as
\[
\frac{\sum_{i\in \mathcal{I}_3} {\Delta_i\mathbf{1}\{R(X_i, T_i)\ge\hat\beta\}}/{\hat{S}_{C\mid X}(T_i\mid X_i)}}{\sum_{i\in \mathcal{I}_3}{\Delta_i}/{\hat{S}_{C\mid X}(T_i\mid X_i)}}
\]
In other words, the empirical coverage is estimated as the weighted proportion of observed failure times that exceed the lower predictive bound, using inverse probability weights derived from the estimated censoring survival function. Similarly, the AIPCW-based coverage computation extends this by incorporating an augmentation term.

The OR-based empirical coverage is computed as $|\mathcal{D}_3|^{-1}\sum_{i\in \mathcal{I}_3} \hat\eta(\hat\beta, 0, X_i)$.
In fact, $\hat\eta(\beta,u\mid X)$ is an estimator of $\eta^\star(\beta,u\mid X)=\mathbb{E}\left[\mathbf{1}\{R(X,T)\ge\beta\} \mid X,T\geq u\right]$. When the non-conformity score is defined through the conditional quantile function of $T \mid X$, the OR-based empirical coverage reduces to $|\mathcal{D}_3|^{-1}\sum_{i\in \mathcal{I}_3} \hat{S}_{T\mid X}(\hat{q}_{T\mid X}(\hat\beta\mid X_i) \mid X_i)$.

Both metrics provide valid but complementary assessments of predictive coverage. By using both evaluation strategies, we obtain a more comprehensive understanding of the calibration and reliability of our methods in this real-world setting.

\subsection{Results}
\begin{figure}
    \centering
    \includegraphics[width=\textwidth]{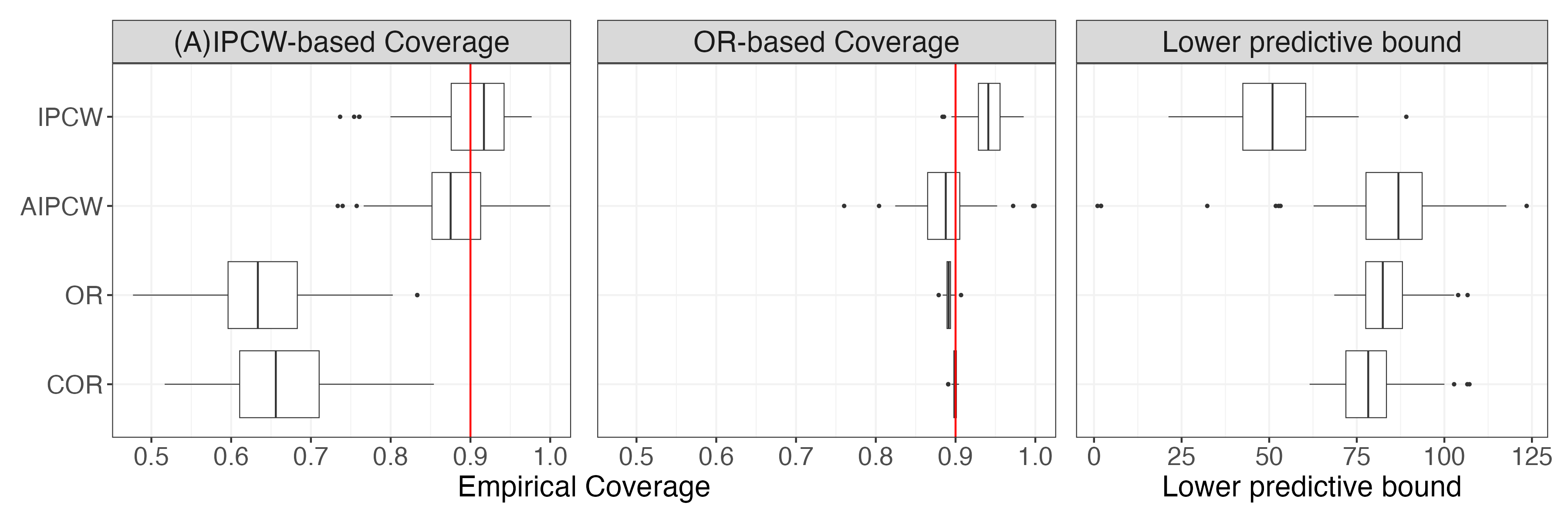}
    \caption{Evaluation on the real-world dataset across 100 random data splits. The left and center panels display the distribution of empirical coverage using (A)IPCW-based  coverage evaluation (left) and OR-based coverage evaluation (center), with red lines marking the 90\% target coverage. The right panel displays the average LPB.
    }
    \label{fig:coverage_boxplot_ra}
\end{figure}

We implement our methods (OR, COR, IPCW, AIPCW) across 100 different random splits of the training, calibration, and test sets, using the Super Learner algorithm. The estimated coefficients of the candidate models used within the SL algorithm, are provided in Section~\ref{App:data} of the Supplementary Material. Figure \ref{fig:coverage_boxplot_ra} presents boxplots of the empirical coverage in the test set for each of these 100 splits.

In the evaluation based on the (A)IPCW metric (left panel), the AIPCW estimator is assessed using the AIPCW-based coverage, while the other three methods are evaluated through IPCW-based coverage. The results indicate that IPCW and AIPCW outperform the regression-based methods, though they exhibit slight deviations from the nominal level, with IPCW tending to overcover and AIPCW slightly undercovering, on average. These deviations are expected given the relatively small sample size, as our theoretical results hold asymptotically. Moreover, considering the low event rate in this dataset, the performance of our methods is strong.

The OR-based evaluation approach (center panel) produces almost perfect coverage for the OR and COR methods. However, this result is a consequence of the fact that the coverage is computed using the same metric employed to estimate the LPBs. Specifically, for the OR method, by construction $\hat{S}_{T\mid X}(\hat{q}_{T\mid X}(\hat\beta\mid X_i)\mid X_i) = 1-\alpha$, since the conditional quantile is directly derived from the estimated survival curve. For COR, we observe a slight deviation from the target level due to the additional calibration step. Meanwhile, the IPCW and AIPCW methods continue to exhibit reasonable coverage even when evaluated using this alternative approach, further reinforcing their reliability.

Additionally, we compute the average estimated LPB values in the test set across the 100 cross-validation splits. The right panel of Figure~\ref{fig:coverage_boxplot_ra} displays the distribution of  these mean LPB estimates for each of the four methods, showing comparable results across methods. IPCW tends to produce smaller LPBs than the others, reflecting a more conservative approach.

\section{Discussion}
In this work, we developed a novel methodology to construct calibrated lower predictive bounds on survival times, addressing key limitations of existing methods. Our approach is flexible and general, accommodating the most common censoring type, whereby either censoring or event time only is observed, whichever comes first, accounting for dependence between failure and censoring times through baseline characteristics, via the use of inverse-probability-of-censoring weighting and augmented-inverse-probability-of-censoring weighting. By leveraging the double robustness property of AIPCW, our method ensures consistency even in the presence of partial model misspecifications, thereby enhancing its reliability in practical applications.

Our simulation studies provided empirical evidence of the effectiveness of the proposed IPCW and AIPCW methodologies. Across a range of scenarios, including settings where standard model assumptions fail, our methods consistently achieved coverage close to the target level, particularly when leveraging nonparametric regression or flexible machine learning methods such as RSF and SL. The comparative analysis also underscored the limitations of naive methods that do not appropriately account for censoring and demonstrated the advantages of incorporating modern semiparametric efficiency theory into predictive frameworks for survival data. Additionally, the outcome regression approaches, OR and COR, proved to be computationally efficient alternatives to IPCW and AIPCW, but less robust due to their heavy dependence on correct model specification.

Despite these contributions, several avenues for future extensions remain open. One promising direction is to extend our framework to handle other coarsened data settings, where the observed data may represent more complex forms of incompleteness or missingness, such as interval censoring. These scenarios often arise in clinical and longitudinal studies and require more nuanced modeling assumptions. By incorporating principles from monotone coarsening and semiparametric efficiency theory, our methods could be adapted to provide valid and efficient predictive bounds in these more general settings.

These extensions would generalize the applicability of our framework, making it a valuable tool for a wider range of real-world applications. Addressing these challenges represents a critical step toward further enhancing the robustness and versatility of predictive methodologies in censored and more general coarsened data frameworks.

%\section*{Acknowledgement}

\bibliographystyle{apalike}
\bibliography{ref}

\newpage
\setcounter{section}{0}
\setcounter{equation}{0}
\setcounter{figure}{0}
\setcounter{algocf}{0}
\setcounter{table}{0}
\renewcommand{\thesection}{S.\arabic{section}}
\renewcommand{\theequation}{E.\arabic{equation}}
\renewcommand{\thefigure}{F.\arabic{figure}}
\renewcommand{\thealgocf}{A.\arabic{algocf}}
\renewcommand{\thetable}{T.\arabic{table}}
\renewcommand{\theHsection}{S.\arabic{section}}
\renewcommand{\theHequation}{E.\arabic{equation}}
\renewcommand{\theHfigure}{F.\arabic{figure}}
\renewcommand{\theHalgocf}{A.\arabic{algocf}}
\renewcommand{\theHtable}{T.\arabic{table}}
% \tableofcontents
% \titlelabel{\thetitle: }
% \cftsetindents{section}{1em}{2.5em}
% \cftsetindents{subsection}{1.5em}{3em}
% \setcounter{page}{1}
  \begin{center}
  \Large {\bf Supplement to ``Doubly Robust and Efficient Calibration of Prediction Sets for Right-Censored Time-to-Event Outcomes''}\label{Supp}
  \end{center}

\begin{abstract}
This supplement includes detailed proofs for all the main results presented in the paper, along with supporting lemmas. Additionally, it contains the algorithms for the proposed procedures in the case of a general non-conformity score,  a summary table of the simulation settings, a description of the competing methods and their implementation, and supplementary simulation results.
\end{abstract}

\section{Proof of equation (\ref{eq:IPCW})}\label{App:1}
This result was originally established by \cite{robins1992recovery} for a general estimating function of the full data; also see \cite{laan2003unified}.
\begin{proof}[Proof of equation (\ref{eq:IPCW})]
\begin{align*}
     \mathbb{P}(R(X,T)\ge \beta) &= \mathbb{E}\left[\mathbf{1}\{R(X,T)\ge \beta\}\right]\\
     &= \mathbb{E}\left[\mathbf{1}\{R(X,T)\ge \beta\}\frac{\mathbb{P}(T\leq C\mid T,X)}{\mathbb{P}(T\leq C\mid T,X)}\right]\\
     &= \mathbb{E}\left[\mathbf{1}\{R(X,T)\ge \beta\}\frac{\mathbb{E}\left[\mathbf{1}\{T\le C\}\mid T,X\right]}{S_{C\mid X}^\star(T\mid X)}\right]\\
     &= \mathbb{E}\left[\mathbb{E}\left[\frac{\mathbf{1}\{R(X,T)\ge \beta\}}{S_{C\mid X}^\star(T\mid X)}\mathbf{1}\{T\leq C\}\bigg\vert T,X\right]\right]\\
     &= \mathbb{E}\left[\frac{\mathbf{1}\{R(X,T)\ge \beta\}}{S_{C\mid X}^\star(T\mid X)}\mathbf{1}\{T\leq C\}\right]\\
     &= \mathbb{E}\left[\frac{\Delta\mathbf{1}\{R(X,T)\ge \beta\}}{S_{C\mid X}^\star(T\mid X)} \right],
\end{align*}
where the third equality follows from the fact that $T \perp \!\!\! \perp C \mid  X$, by Assumption \ref{Assum:CIC}.
\end{proof}
%Importantly, the proof remains unchanged if we replace $\mathbf{1}\{T\ge \hat L(X)\}$  with any function of $(X, T)$.  
%This result was also extended in even more general coarsening scenarios. E.g. \cite{tsiatis2007semiparametric} does the inverse probability weighted complete-case estimator under monotone coarsening and coarsening at random. This last assumption is a generalization of conditionally independent censoring. 
The expression on the RHS of (\ref{eq:IPCW}) depends solely on the observed data. In fact, since all terms are multiplied by $\Delta$, we can rewrite the dependence on $T$ as dependence on $Y$.
%\[
%\mathbb{E}\left[\frac{\Delta\mathbf{1}\{Y\ge \hat L(X)\}}{S_{C\mid X}^\star(Y\mid X)} \right].
%\]

\section{Proof of Theorem \ref{thm:valid-IPCW}}\label{App:2}
We begin by stating and proving two key lemmas, which serve as the primary steps toward establishing Theorem \ref{thm:valid-IPCW}.

\begin{lemma}\label{lem:bound-Op} 
Under assumptions \ref{Assum:CIC} and \ref{Assum:bound}, for any $\epsilon>0$,
\begin{equation}\label{eq:bound-Op}
\begin{aligned}
&\sup_{\beta\in[0, 1]}\,\left|\frac{1}{\left| \mathcal{D}_2\right| } \sum_{i \in \mathcal{I}_2} \frac{\Delta_i \left\{\mathbf{1}\{R(X_i,T_i\} \ge \beta) - (1-\alpha)\right\}}{\hat{S}_{C\mid X}(T_i\mid X_i)} - \mathbb{E}\left[\frac{\Delta \left\{\mathbf{1}\{R(X,T)\ge\beta\}-(1-\alpha)\right\}}{\hat{S}_{C\mid X}(T\mid X)}\mid \mathcal{D}\right]\right| \\
&\quad \le s_0\left(\left(\frac{1}{2}\log\frac{1}{\epsilon}\right)^{1/2}+K\right)\frac{1}{\sqrt{\left| \mathcal{D}_2\right| }},
\end{aligned}
\end{equation}
with probability at least $1-\epsilon$, for some constant $K$.
\end{lemma}
\begin{proof}
Let $n=\left| \mathcal{D}_2\right| $. Let $f:\mathbb{R}^{d+2}\rightarrow\mathbb{R}$ be defined as
\[
f(x,\delta,t)=\frac{\delta\left\{\mathbf{1}\{R(x,t)\ge \beta\}-(1-\alpha)\right\}}{\hat{S}_{C\mid X}(t\mid x)},
\]
given $\beta, R, \hat{S}_{C\mid X}$ fixed. Note that $f$ is only a function of $(x,\delta,\delta t)$ since there is a multiplicative factor of $\delta$. 
Define the empirical process
\[
\mathbb G_n f := \frac{1}{\sqrt{n}} \sum_{i=1}^n\left(f(X_i,\Delta_i,T_i)-\mathbb{E}\left[f(X,\Delta,T)\mid \mathcal{D}\right]\right).
\]
We want to show that for any $\epsilon> 0$ there exists $M(\epsilon)$ such that
\begin{equation*}
\mathbb P\left(\sup_{\beta\in[0, 1]}\left|\mathbb G_n f \right| \ge M(\epsilon)\right)\le \epsilon.
\end{equation*}
We apply the well-known McDiarmid's inequality to the function $\sup_\beta\mid \mathbb{G}_n f\mid $. We first verify the bounded differences property, which is necessary for McDiarmid's inequality to hold. Given $j\in[n]$ fixed,
\begin{align*}
    &\sup_{\substack{(x_1,\delta_1,t_1),\dots, (x_n,\delta_n,t_n) \\  (x_j',\delta_j',t_j')}}\Bigg|\sup_{\beta\in[0, 1]}\left|\frac{1}{\sqrt{n}} \sum_{i=1}^n\left(f(x_i,\delta_i,t_i) - \mathbb{E}\left[f(X, \Delta,T)\mid\mathcal{D}\right]\right)\right| \\
    &\quad -\sup_{\beta\in[0, 1]}\Bigg|\frac{1}{\sqrt{n}} \sum_{i\neq j} \left(f(x_i,\delta_i,t_i) - \mathbb{E}\left[f(X,\Delta,T)\mid\mathcal{D}\right]\right) \\
    &\quad +\frac{1}{\sqrt{n}} \left(f(x_j',\delta_j',t_j') - \mathbb{E}\left[f(X,\Delta,T)\mid\mathcal{D}\right]\right)\Bigg|\Bigg|\\
    &\le \sup_{\substack{(x_1,\delta_1,t_1),\dots, (x_n,\delta_n,t_n) \\  (x_j',\delta_j',t_j')}}\sup_{\beta\in[0, 1]} \Bigg| \frac{1}{\sqrt{n}} \sum_{i=1}^n f(x_i,\delta_i,t_i)-\frac{1}{\sqrt{n}} \sum_{i\neq j} f(x_i,\delta_i,t_i) - \frac{1}{\sqrt{n}} f(x_j',\delta_j',t_j') \Bigg|\\
    &= \sup_{\substack{(x_1,\delta_1,t_1),\dots, (x_n,\delta_n,t_n) \\  (x_j',\delta_j',t_j')}}\sup_{\beta\in[0, 1]} \frac{1}{\sqrt{n}} \Bigg| \frac{\delta_j\left\{\mathbf{1}\{R(x_j,t_j)\ge\beta\}-(1-\alpha)\right\}}{\hat{S}_{C\mid X}(t_j\mid x_j)} \\
    &\quad- \frac{\delta_j'\left\{\mathbf{1}\{R(x_j',t_j')\beta\}-(1-\alpha)\right\}}{\hat{S}_{C\mid X}(t_j';x_j')} \Bigg|\\
    &< \frac{s_0}{\sqrt{n}},
\end{align*}
where the first inequality follows by applying $\sup f - \sup g \le \sup(f-g)$, then $\left| \sup f\right| \le\sup \left| f\right| $, then $\left| \left| a\right| -\left| b\right| \right| \le \left| a-b\right|$, and the expected values get simplified; the second inequality follows by applying Assumption \ref{Assum:bound} and by bounding $\delta$ and the indicator by 1. Therefore, we can apply McDiarmid's inequality which yields
\[
\mathbb P\left(\sup_{\beta\in[0, 1]}\left| \mathbb G_n f \right|  - \mathbb{E}\left[\sup_{\beta\in[0, 1]}\left| \mathbb G_n f \right| \right] \ge v\right)\le \exp\left(-\frac{2v^2}{\sum_{i=1}^n s_0^2/n}\right)=\exp\left(-\frac{2v^2}{s_0^2}\right),
\]
for any $v>0$.
Moreover, by Lemma 4 of \cite{yang2024doubly}, there exists a universal constant $K$ such that
%with $h_1(x,\delta,t)=\frac{\delta}{\widehat{S}_{C\mid X}(t\mid x)}$ and $h_2(t,x)=\widehat{R}(t,x)$ we have
\[
\mathbb{E}\left[\sup_{\beta\in[0, 1]}\left| \mathbb G_n f \right| \right]\le Ks_0.
\]
Then, for any $v>0$
\begin{align*}
    \mathbb P\left(\sup_{\beta\in[0, 1]}\left| \mathbb G_n f \right|  \ge v + Ks_0\right)&\le \mathbb P\left(\sup_{\beta\in[0, 1]}\left| \mathbb G_n f \right|  - \mathbb{E}\left[\sup_{\beta\in[0, 1]}\left| \mathbb G_n f \right| \right] \ge v\right)\\
    &\le \exp\left(-\frac{2v^2}{s_0^2}\right).
\end{align*}
Solving for $v$ in $\epsilon = \exp\left(-\frac{2v^2}{s_0^2}\right)$, gives $v=s_0\left(\frac{1}{2}\log\frac{1}{\epsilon}\right)^{1/2}$. Hence, for any $\epsilon>0$
\[
\mathbb P\left(\sup_{\beta\in[0, 1]}\left| \mathbb G_n f \right|  \ge s_0\left(\left(\frac{1}{2}\log\frac{1}{\epsilon}\right)^{1/2}+K\right) \right) \le \epsilon.
\]
\end{proof}

\begin{lemma}\label{lem:bound-norm}
Under assumptions \ref{Assum:CIC} and \ref{Assum:bound},
\begin{align}\label{eq:bound-norm}
&\sup_{\beta\in[0, 1]} 
\left|\mathbb{E}\left[\frac{\Delta\left\{\mathbf{1}\{R(X,T)\ge\beta\}-(1-\alpha)\right\}}{\hat{S}_{C\mid X}(T\mid X)}\mid \mathcal{D}\right] \right. - \left.\mathbb{E}\left[\frac{\Delta\left\{\mathbf{1}\{R(X,T)\ge\beta\}-(1-\alpha)\right\}}{S_{C\mid X}^\star(T\mid X)}\mid \mathcal{D}\right]\right| \notag \\
&\quad < s_0^2 \left\| \hat{S}_{C\mid X} - S_{C\mid X}^\star\right\| _{L^2}.
\end{align}
\end{lemma}
\begin{proof}
For any fixed $\beta\in[0, 1]$ we have
\begin{align*}
    &\left|\mathbb{E}\left[\frac{\Delta\left\{\mathbf{1}\{R(X,T)\ge\beta\}-(1-\alpha)\right\}}{\hat{S}_{C\mid X}(T\mid X)}\mid \mathcal{D}\right] - \mathbb{E}\left[\frac{\Delta\left\{\mathbf{1}\{R(X,T)\ge\beta\}-(1-\alpha)\right\}}{S_{C\mid X}^\star(T\mid X)}\mid \mathcal{D}\right]\right|\\
    &=\left|\mathbb{E}\left[\Delta\left\{\mathbf{1}\{R(X,T)\ge\beta\}-(1-\alpha)\right\}\left(\frac{1}{\hat{S}_{C\mid X}(T\mid X)}- \frac{1}{{S_{C\mid X}^\star(T\mid X)}} \right)\mid \mathcal{D}\right] \right|\\
    &\le \mathbb{E}\left[\left|\Delta\left\{\mathbf{1}\{R(X,T)\ge\beta\}-(1-\alpha)\right\}\left(\frac{1}{\hat{S}_{C\mid X}(T\mid X)}- \frac{1}{{S_{C\mid X}^\star(T\mid X)}} \right)\right|\mid \mathcal{D} \right]\\
    &\le \mathbb{E}\left[\left|\frac{1}{\hat{S}_{C\mid X}(T\mid X)}- \frac{1}{{S_{C\mid X}^\star(T\mid X)}}\right|\mid \mathcal{D}\right]\\
    &=\mathbb{E}\left[\frac{\left|\hat{S}_{C\mid X}(T\mid X)-S_{C\mid X}^\star(T\mid X)\right|}{\widehat{S}_{C\mid X}(T\mid X)S_{C\mid X}^\star(T\mid X)}\mid \mathcal{D} \right]\\
    &\le \sqrt{\mathbb{E}\left[\left(\hat{S}_{C\mid X}(T\mid X)-S_{C\mid X}^\star(T\mid X)\right)^2\mid \mathcal{D}\right]\mathbb{E}\left[\frac{1}{\hat{S}_{C\mid X}(T\mid X)^2S_{C\mid X}^\star(T\mid X)^2}\mid \mathcal{D}\right]}\\
    &=\left\Vert\hat{S}_{C\mid X}-S_{C\mid X}^\star\right\Vert_{L^2}\sqrt{\mathbb{E}\left[\frac{1}{\hat{S}_{C\mid X}(T\mid X)^2S_{C\mid X}^\star(T\mid X)^2}\mid \mathcal{D}\right]}\\
    &\le s_0^2\left\Vert\hat{S}_{C\mid X}-S_{C\mid X}^\star\right\Vert_{L^2},
\end{align*}
where the first inequality follows by Jensen's; the second inequality follows because
$\left|\Delta\left\{\mathbf{1}\{R(X,T)\ge\beta\}-(1-\alpha)\right\}\right| \le 1 \quad \mathrm{a.s.}$;
the third inequality follows by Cauchy-Schwarz; the last inequality follows by assumption \ref{Assum:bound}. Taking the supremum over $\beta\in[0,1]$, we get (\ref{eq:bound-norm}).
\end{proof}

\begin{proof}[Proof of Theorem \ref{thm:valid-IPCW}]
We first prove the marginal coverage guarantee. Let $n=\left| \mathcal{D}_2\right| $. Fix $\epsilon\in(0,1)$. For any fixed $\beta\in[0,1]$,
\begin{align*}
    &\left|\frac{1}{n}\sum_{i=1}^n \frac{\Delta_i\left\{\mathbf{1}\{R(X_i,T_i)\ge\beta\}-(1-\alpha)\right\}}{\hat{S}_{C\mid X}(T_i\mid X_i)} - \left\{\mathbb{P}\left(R(X,T)\ge\beta\mid \mathcal{D}\right)-(1-\alpha)\right\} \right|\\
    &\overset{(\ref{eq:IPCW})}=\left|\frac{1}{n}\sum_{i=1}^n \frac{\Delta_i\left\{\mathbf{1}\{R(X_i,T_i)\ge\beta\}-(1-\alpha)\right\}}{\hat{S}_{C\mid X}(T_i\mid X_i)} -\mathbb{E}\left[\frac{\Delta\mathbf{1}\{R(X,T)\ge\beta\}}{S_{C\mid X}^\star(T\mid X)}\mid \mathcal{D}\right]+(1-\alpha)\right|\\
    &=\Bigg|\frac{1}{n}\sum_{i=1}^n \frac{\Delta_i\left\{\mathbf{1}\{R(X_i,T_i)\ge\beta\}-(1-\alpha)\right\}}{\hat{S}_{C\mid X}(T_i\mid X_i)} -\mathbb{E}\left[\frac{\Delta\mathbf{1}\{R(X,T)\ge\beta\}}{S_{C\mid X}^\star(T\mid X)}\mid \mathcal{D}\right]\\
    &\quad +(1-\alpha)\mathbb{E}\left[\frac{\Delta}{S_{C\mid X}^\star(T\mid X)}\mid \mathcal{D}\right]\Bigg|\\
    &=\left|\frac{1}{n}\sum_{i=1}^n \frac{\Delta_i\left\{\mathbf{1}\{R(X_i,T_i)\ge\beta\}-(1-\alpha)\right\}}{\hat{S}_{C\mid X}(T_i\mid X_i)} -\mathbb{E}\left[\frac{\Delta\left\{\mathbf{1}\{R(X,T)\ge\beta\}-(1-\alpha)\right\}}{S_{C\mid X}^\star(T\mid X)}\mid \mathcal{D}\right]\right|\\
    &\le\sup_{\gamma\in[0, 1]}\left|\frac{1}{n}\sum_{i=1}^n\frac{\Delta_i\left\{\mathbf{1}\{R(X_i,T_i)\ge\gamma\}-(1-\alpha)\right\}}{\hat{S}_{C\mid X}(T_i\mid X_i)}-\mathbb{E}\left[\frac{\Delta\left\{\mathbf{1}\{R(X,T)\ge\gamma\}-(1-\alpha)\right\}}{S_{C\mid X}^\star(T\mid X)}\mid \mathcal{D}\right]\right|\\
    &\le \sup_{\gamma\in[0, 1]}\,\left|\frac{1}{n}\sum_{i=1}^n\frac{\Delta_i\left\{\mathbf{1}\{R(X_i,T_i)\ge\gamma\}-(1-\alpha)\right\}}{\hat{S}_{C\mid X}(T_i\mid X_i)}-\mathbb{E}\left[\frac{\Delta\left\{\mathbf{1}\{R(X,T)\ge\gamma\}-(1-\alpha)\right\}}{\hat{S}_{C\mid X}(T\mid X)}\mid \mathcal{D}\right]\right|\\
    &\quad + \sup_{\gamma\in[0, 1]}\left|\mathbb{E}\left[\frac{\Delta\left\{\mathbf{1}\{R(X,T)\ge\gamma\}-(1-\alpha)\right\}}{\hat{S}_{C\mid X}(T\mid X)}\mid \mathcal{D}\right] - \mathbb{E}\left[\frac{\Delta\left\{\mathbf{1}\{R(X,T)\ge\gamma\}-(1-\alpha)\right\}}{S_{C\mid X}^\star(T\mid X)}\mid \mathcal{D}\right]\right|\\
    &\stackrel{\mathrm{\eqref{eq:bound-Op},\eqref{eq:bound-norm}}}\le s_0\left(\left(\frac{1}{2}\log\frac{1}{\epsilon}\right)^{1/2}+K\right)\frac{1}{\sqrt{n }}+s_0^2\| \hat{S}_{C\mid X}-S_{C\mid X}^\star\| _{L^2},
\end{align*}
with probability at least $1-\epsilon$. Letting $\beta=\hat{\beta}_{\mathrm{IPCW}}$ and rearranging the inequality above, we get
\begin{align*}
    \mathbb{P}\left(R(X,T)\ge\hat{\beta}_{\mathrm{IPCW}}\mid \mathcal{D}\right) - (1-\alpha) &> \frac{1}{n}\sum_{i=1}^n \frac{\Delta_i\left\{\mathbf{1}\{R(X_i,T_i)\ge\hat{\beta}_{\mathrm{IPCW}}\}-(1-\alpha)\right\}}{\hat{S}_{C\mid X}(T_i\mid X_i)}\\
    &\quad-s_0\left(\left(\frac{1}{2}\log\frac{1}{\epsilon}\right)^{1/2}+K\right)\frac{1}{\sqrt{\left| \mathcal{D}_2\right| }}-s_0^2\| \hat{S}_{C\mid X} - S_{C\mid X}^\star\| _{L^2}\\
    &\ge -s_0\left(\left(\frac{1}{2}\log\frac{1}{\epsilon}\right)^{1/2}+K\right)\frac{1}{\sqrt{\left| \mathcal{D}_2\right| }}-s_0^2\| \hat{S}_{C\mid X}-S_{C\mid X}^\star\|_{L^2},
\end{align*}
with probability at least $1-\epsilon$, where the last inequality follows because, by construction, $\hat{\beta}_{\mathrm{IPCW}}$ satisfies
\[
\frac{1}{\left| \mathcal{D}_2\right| }\sum_{i\in\mathcal{I}_2}\frac{\Delta_i\left\{\mathbf{1}\{R(X_i,T_i)\ge\hat{\beta}_{\mathrm{IPCW}}\}-(1-\alpha)\right\}}{\hat{S}_{C\mid X}(T_i\mid X_i)} \ge 0.
\]

We now prove the conditional coverage guarantee.
For any $\beta\in[0,1]$,
\begin{align*}
    \mathbb{P}\left(T\ge \hat{q}_{T\mid X}(\beta\mid X) \mid \mathcal{D}\right) &=\mathbb{E}\left[S_{T\mid X}^\star\left(\hat{q}_{T\mid X}(\beta\mid X)\mid X\right)\mid \mathcal{D}\right]\\
    &= \mathbb{E}\left[S_{T\mid X}^\star\left(\hat{q}_{T\mid X}(\beta\mid X)\mid X\right)+ 1-\beta-\hat{S}_{T\mid X}\left(\hat{q}_{T\mid X}(\beta\mid X)\mid X\right)\mid \mathcal{D}\right] \\
    &\le 1-\beta + \|\hat{S}_{T\mid X} - S_{T\mid X}^\star\|_\infty.
\end{align*}
Let 
\[
\hat{W}(\beta) = \frac{1}{n} \sum_{i \in \mathcal{I}_2} \frac{\Delta_i \left\{\mathbf{1}\{T_i \ge \hat{q}_{T\mid X}(\beta\mid X_i)\} - (1-\alpha)\right\}}{\hat{S}_{C\mid X}(T_i \mid  X_i)},
\]
and let 
\[
\gamma=\|\hat{S}_{T\mid X} - S_{T\mid X}^\star\|_\infty+ s_0\left(\left(\frac{1}{2}\log\frac{1}{\epsilon}\right)^{1/2}+K\right)\frac{1}{\sqrt{n }}+s_0^2\| \hat{S}_{C\mid X}-S_{C\mid X}^\star\|_{L^2}.
\]
As showed in the first part of the proof, by~\eqref{eq:bound-Op},~\eqref{eq:bound-norm}, 
\begin{align*}
    \hat{W}(\alpha+\gamma) &< \mathbb{P}\left(T\ge \hat{q}_{T\mid X}(\alpha+\gamma\mid X) \mid \mathcal{D}\right) - (1-\alpha)\\
    &\quad + s_0\left(\left(\frac{1}{2}\log\frac{1}{\epsilon}\right)^{1/2}+K\right)\frac{1}{\sqrt{n }}+s_0^2\| \hat{S}_{C\mid X}-S_{C\mid X}^\star\|_{L^2}\\
    &\le 1-\alpha-\gamma + \|\hat{S}_{T\mid X} - S_{T\mid X}^\star\|_\infty -(1-\alpha)\\
    &\quad + s_0\left(\left(\frac{1}{2}\log\frac{1}{\epsilon}\right)^{1/2}+K\right)\frac{1}{\sqrt{n }}+s_0^2\| \hat{S}_{C\mid X}-S_{C\mid X}^\star\|_{L^2},\\
    &=0,
\end{align*}
with probability at least $1-\epsilon$ over $\mathcal{D}$. Similarly, we can show that $\hat{W}(\alpha-\gamma) >0$, with probability at least $1-\epsilon$ . And since $\hat{\beta}_{\mathrm{IPCW}}= \sup\{\beta\in[0,1]:\hat{W}(\beta)\ge 0\}$, we get that $|\hat{\beta}_{\mathrm{IPCW}}-\alpha|\le\gamma$, with probability at least $1-\epsilon$. 

Note that
\begin{align*}
&\mathbb{P}(T\ge\hat{q}_{T|X}(\hat{\beta}_{IPCW}|X)|X,\mathcal{D})-(1-\alpha) \\
&=\mathbb{P}(\hat{S}_{T|X}(T|X)\le1-\hat{\beta}_{IPCW}|X,\mathcal{D})-(1-\alpha) \\
&\le\mathbb{P}(S_{T|X}^\star(T|X)\le1-\alpha+\|\hat{S}_{T|X}-S_{T|X}^\star\|_{\infty}+|\hat{\beta}_{IPCW}-\alpha\|X,\mathcal{D})-(1-\alpha).
\end{align*}
Similarly,
\begin{align*}
&\mathbb{P}(T\ge\hat{q}_{T|X}(\hat{\beta}_{IPCW}|X)|X,\mathcal{D})-(1-\alpha) \\
&\ge\mathbb{P}(S_{T|X}^\star(T|X)\le1-\alpha-\|\hat{S}_{T|X}-S_{T|X}^\star\|_{\infty}-|\hat{\beta}_{IPCW}-\alpha\|X,\mathcal{D})-(1-\alpha).
\end{align*}
Note that $S_{T|X}^\star(T|X)$ is independent of $X$ and is uniformly distributed (under the assumption of continuity
of the distribution of $T\mid X$). Hence, we get
$$|\mathbb{P}(T\ge\hat{q}_{T|X}(\hat{\beta}_{IPCW}|X)|X,\mathcal{D})-(1-\alpha)| \le |\hat{\beta}_{IPCW}-\alpha| + \|\hat{S}_{T|X}-S_{T|X}^\star\|_{\infty}.$$
Therefore, with probability at least $1-\epsilon,$ we have
\begin{align*}
    &|\mathbb{P}(T\ge\hat{q}_{T|X}(\hat{\beta}_{IPCW}|X)|X,\mathcal{D})-(1-\alpha)|\\
    &\le s_0^2\|\hat{S}_{C|X}-S_{C|X}^\star\|_{L^{2}}+2\|\hat{S}_{T|X}-S_{T|X}^\star\|_{\infty}+s_0\left(\left(\frac{1}{2}\log\frac{1}{\epsilon}\right)^{1/2}+K\right)\frac{1}{\sqrt{n}}.
\end{align*}
\end{proof}

\section{Proof of the Orthogonal Projection formula for AIPCW}\label{App:aipcw-projection}
We provide a self-contained derivation for the orthogonal projection of the IPCW estimating function $W(\beta)$ onto the censoring tangent space $\mathcal{T}$. This result was originally established by \cite{robins1992recovery}; also see \citet[Sections 9.3, 10.4]{tsiatis2007semiparametric}. 

\begin{proof}
We want to show that the orthogonal projection of $W$ onto the censoring tangent space is given by
\[
\Pi\left(W(\beta)\mid \mathcal{T}\right)=-\int \frac{\eta^\star(\beta,u|X)- (1-\alpha) }{S_{C\mid X}^\star\left(u\mid X\right)} dM_{C\mid X}^\star\left( u\right). 
\]
To confirm the proposed formula, we show that the resulting influence function is orthogonal to every element $h^\star \in \mathcal{T}$. That is, for an arbitrary $h^\star = \int h(u,X)dM_{C|X}^\star(u|X) \in \mathcal{T}$, it must hold that $\mathbb{E}\left[\left(W(\beta) + \Pi(W(\beta)|\mathcal{T})\right) h^\star\right] = 0$. I.e., 
\begin{align}\label{eq:proj1}
    \mathbb{E}[W(\beta) h^\star] &= -\mathbb{E}[\Pi(W(\beta)|\mathcal{T}) h^\star]\notag \\
    &= -\mathbb{E}\left[\int\frac{\eta^\star(\beta,u|X)-(1-\alpha)}{S_{C|X}^\star(u|X)}dM_{C|X}^\star(u|X) \int h(v,X)dM_{C|X}^\star(v|X)\right]\notag \\
    &= -\mathbb{E}\left[\int \frac{\eta^{*}(\beta,u|X)-(1-\alpha)}{S_{C|X}^\star(u|X)} h(u,X) S_{C|X}^\star(u|X) d\Lambda_{C|X}^\star(u|X)\right]\notag \\
    &= -\mathbb{E}\left[\int \{\eta^\star(\beta,u|X)-(1-\alpha)\} h(u,X) d\Lambda_{C|X}^\star(u|X)\right], 
\end{align}
where in the third equality we used the the Martingale central limit theorem for the cross-product of two martingale difference integrals.

On the other end, substituting the definitions of $W(\beta)$ and $h^\star$, we get
\begin{align*}
    \mathbb{E}[W(\beta) h^\star] &= \mathbb{E}\left[\frac{\Delta(\mathbf{1}\{R(X,T)\ge\beta\}-(1-\alpha))}{S_{C|X}^\star(T|X)} \int h(u,X)dM_{C|X}^\star(u|X)\right]\\
    &=\mathbb{E}\left[(\mathbf{1}\{R(X,T)\ge\beta\}-(1-\alpha))  \mathbb{E}\left[\frac{\Delta}{S_{C|X}^\star(T|X)} \int h(u,X)dM_{C|X}^{*}(u|X) \mid X, T \right]\right]\\
    &= -\mathbb{E}\left[(\mathbf{1}\{R(X,T)\ge\beta\}-(1-\alpha)) \int \mathbf{1}\{T \ge u\} h(u,X) d\Lambda_{C|X}^\star(u|X)\right]\\
    &= -\mathbb{E}\left[\int \mathbb{E}\left[(\mathbf{1}\{R(X,T)\ge\beta\}-(1-\alpha)) \mathbf{1}\{T \ge u\} \mid X\right] h(u,X) d\Lambda_{C|X}^\star(u|X)\right]
\end{align*}
where the third equality follows by the decomposition of the martingale difference process, and the fourth by Fubini's theorem. The inner conditional expectation is
\begin{align*}
&\mathbb{E}[(\mathbf{1}\{R(X,T)\ge\beta\}-(1-\alpha)) \mathbf{1}\{T \ge u\} \mid X] \\
&= \mathbb{E}[\mathbf{1}\{R(X,T)\ge\beta\} 1\{T \ge u\} | X] - (1-\alpha) \mathbb{E}[\mathbf{1}\{T \ge u\} \mid X] \\
&= \mathbb{E}[\mathbf{1}\{R(X,T)\ge\beta\} | X, T \ge u] \mathbb{P}(T \ge u|X) - (1-\alpha) \mathbb{P}(T \ge u|X) \\
&= \eta^\star(\beta,u|X) S_{T|X}^\star(u|X) - (1-\alpha) S_{T|X}^\star(u|X) \\
&= \{\eta^\star(\beta,u|X) - (1-\alpha)\} S_{T|X}^\star(u|X)
\end{align*}
Substituting this back we get
\begin{equation}\label{eq:proj2}
    \mathbb{E}[W(\beta) h^*] = -\mathbb{E}\left[\int \{\eta^{*}(\beta,u|X)-(1-\alpha)\} S_{T|X}^\star(u|X) h(u, X) d\Lambda_{C|X}^\star(u|X)\right]. 
\end{equation}
Comparing~\eqref{eq:proj1} and~\eqref{eq:proj2}, the terms are exactly equal, which is what we needed to prove. Therefore, $\mathbb{E}[(W(\beta) + \Pi(W(\beta)|\mathcal{T})) h^\star] = 0$, confirming that $W(\beta) + \Pi(W(\beta)|\mathcal{T})$ is orthogonal to the censoring tangent space $\mathcal{T}$.
\end{proof}

\section{Proof of Lemma \ref{lem:IF}}\label{App:3}

\begin{proof}[Proof of Lemma \ref{lem:IF}]
To prove the result we will use the two following identities
\begin{equation}\label{eq:SC}
    \int\frac{
    d\mathbf{1}\{Y\in du,\Delta =0\}}{S_{C\mid X}\left( u\mid X\right) } = \frac{1-\Delta}{S_{C\mid X}\left( C\mid X\right) };
\end{equation}
\begin{align}\label{eq:SCT}
    \int\frac{\mathbf{1}\{Y \ge u\} \, d\Lambda_{C\mid X}\left( u \mid  X \right)}{
    S_{C\mid X}\left( u \mid  X \right) }&= 
    \int_{0}^{Y}\frac{\lambda_{C\mid X}\left( u \mid  X \right)}{
    S_{C\mid X}\left( u \mid  X \right) } \, du \nonumber \\
    &= \int_{0}^{T\wedge C}\frac{\lambda_{C\mid X}\left( u \mid  X \right)}{
    S_{C\mid X}\left( u \mid  X \right) } \, du \nonumber \\
    &= \Delta \int_{0}^{T}\frac{
    \lambda_{C\mid X}\left( u \mid  X \right) }{S_{C\mid X}\left( u \mid  X \right) } \, du + (1 - \Delta) \int_{0}^{C}\frac{\lambda_{C\mid X}\left( u \mid  X \right) }{
    S_{C\mid X}\left( u \mid  X \right) } \, du \nonumber \\
    &= \Delta \int_{0}^{T}\frac{\partial }{\partial u}\left( 
    \frac{1}{S_{C\mid X}\left( u \mid  X \right) }\right) du \nonumber \\
    &\quad+(1 - \Delta) \int_{0}^{C}\frac{\partial }{\partial u}\left( \frac{1%
    }{S_{C\mid X}\left( u \mid  X \right) }\right) du \nonumber \\
    &= -\Delta \left[ 1 - \frac{1}{S_{C\mid X}\left( T \mid  X \right) }\right] - (1 - \Delta) \left[ 1 - \frac{1}{S_{C\mid X}\left( C \mid  X \right) }\right].
\end{align}
Combining equations (\ref{eq:SC}) and (\ref{eq:SCT}) yields
\begin{align}\label{eq:M}
\frac{\Delta}{S_{C\mid X}\left( T \mid  X \right)} - 1
&= -\Delta 
\left[ 1 - \frac{1}{S_{C\mid X}\left( T \mid  X \right)} \right] - (1 - \Delta) \nonumber \\
&= -\Delta 
\left[ 1 - \frac{1}{S_{C\mid X}\left( T \mid  X \right)} \right] - \frac{1 - \Delta}{S_{C\mid X}\left( C \mid  X \right)} - (1 - \Delta) 
\left[ 1 - \frac{1}{S_{C\mid X}\left( C \mid  X \right)} \right] \nonumber \\
&= -\int \frac{d\mathbf{1}\{Y \in du, \Delta = 0 \}}{S_{C\mid X}\left( u \mid  X \right)}
+ \int \frac{\mathbf{1}\{Y \ge u\} \, d\Lambda_{C\mid X}\left( u \mid  X \right)}{S_{C\mid X}\left( u \mid  X \right)} \, du \nonumber \\
&= -\int \frac{dM_{C\mid X}\left( u \mid X \right)}{S_{C\mid X}\left( u \mid  X \right)}.
\end{align}
Therefore,
\begin{align*}
& \frac{\Delta}{S_{C\mid X}\left( T\mid X\right) }\left\{
\mathbf{1}\{R(X,T)\ge\beta\} -(1-\alpha) \right\} +\int \frac{\eta(\beta,u\mid X) -(1-\alpha)}{S_{C\mid X}\left(u\mid X\right)}dM_{C\mid X}\left( u\mid X\right) \\
&= \left\{ \frac{\Delta }{S_{C\mid X}\left( T\mid X\right) } -1\right\} \left\{
\mathbf{1}\{R(X,T)\ge\beta\} -(1-\alpha) \right\} + \mathbf{1}\{R(X,T)\ge\beta\}-(1-\alpha) \\
&\quad +\int 
\frac{\eta(\beta,u\mid X) -(1-\alpha) }{
S_{C\mid X}\left(u\mid X\right)}dM_{C\mid X}\left( u\mid X\right) \\
&\overset{\mathrm{(\ref{eq:M})}}= -\int \frac{dM_{C\mid X}\left( u\mid X\right) }{S_{C\mid X}\left( u\mid X\right) }
\left\{ \mathbf{1}\{R(X,T)\ge\beta\} -(1-\alpha)\right\} + \mathbf{1}\{R(X,T)\ge\beta\}-(1-\alpha) \\
&\quad+\int \frac{\eta(\beta,u\mid X)-(1-\alpha) }{S_{C\mid X}\left(
u\mid X\right) }dM_{C\mid X}\left( u\right) \\
&= -\int \frac{dM_{C\mid X}\left( u\mid X\right) }{S_{C\mid X}\left( u\mid X\right) }\left\{
\mathbf{1}\{R(X,T)\ge\beta\} -\eta(\beta,u\mid X) 
\right\}+ \mathbf{1}\{R(X,T)\ge\beta\}-(1-\alpha). 
\end{align*}
\end{proof}

\section{Proof of Theorem \ref{thm:dbl_rob}}\label{App:4}
\begin{proof}[Proofof Theorem \ref{thm:dbl_rob}]
Consider a fixed $\beta$ in $[0,1]$. Then,
\begin{align*}
    &\left|\mathbb{E}\left[\mathrm{IF}_\beta(\hat S_{C\mid X}, \hat\eta)\right]-\mathbb{E}\left[\mathrm{IF}_\beta(S_{C\mid X}^\star,\eta^\star)\right]\right|\\
    &= \bigg|\mathbb{E}\left[\mathbf{1}\{R(X,T)\ge\beta\}-(1-\alpha) -\int \left\{\mathbf{1}\{R(X,T)\ge\beta\}-\hat\eta(\beta,u\mid X) \right\}\frac{d\hat M_{C\mid X}(u\mid X)}{\hat S_{C\mid X}(u\mid X)}\right]\\
    &\quad - \mathbb{E}\left[\mathbf{1}\{R(X,T)\ge\beta\}-(1-\alpha) -\int \left\{\mathbf{1}\{R(X,T)\ge\beta\}-\eta^\star(\beta,u\mid X) \right\}\frac{dM_{C\mid X}^\star(u\mid X)}{S_{C\mid X}^\star(u\mid X)}\right]\bigg|\\
    &= \left|\mathbb{E}\left[\mathbb{E}\left[\int \left\{\mathbf{1}\{R(X,T)\ge\beta\}-\hat\eta(\beta,u\mid X) \right\}\frac{d\hat M_{C\mid X}(u\mid X)}{\hat S_{C\mid X}(u\mid X)}\right]\mid X\right]\right|\\
    &=\left|\mathbb{E}\left[\mathbb{E}\left[\int \left\{\mathbf{1}\{R(X,T)\ge\beta\} -\hat\eta\left(\beta,u\mid X\right) \right\} \frac{dN_{C\mid X}(u\mid X) - \mathbf{1}\{Y\ge u\} d\hat\Lambda_{C\mid X}\left( u\mid X\right)}{\hat{S}_{C\mid X}\left(u\mid X\right) }\mid X\right]\right]\right|\\
    &=\Bigg|\mathbb{E}\Bigg[\mathbb{E}\Bigg[\int \frac{\mathbf{1}\{R(X,T)\ge\beta\} -\hat\eta\left(\beta,u\mid X\right)}{\hat{S}_{C\mid X}\left(u\mid X\right) } \mathbb{E}\big[dN_{C\mid X}(u\mid X) \\
    &\quad\quad -\mathbf{1}\{Y\ge u\}d\hat\Lambda_{C\mid X}\left( u\mid X\right)\mid X,T\big]\mid X\Bigg]\Bigg]\Bigg|\\
    &=\Bigg|\mathbb{E}\Bigg[\int\mathbb{E}\bigg[ \frac{\mathbf{1}\{R(X,T)\ge\beta\} -\hat\eta(\beta,u\mid X)}{\hat{S}_{C\mid X}\left(u\mid X\right)}\mathbb{E}\Big[\mathbb{E}\left[dN_{C\mid X}(u\mid X)\mid C\ge u, X,T\right]\\
    &\quad\quad -\mathbf{1}\{Y\ge u\}d\hat\Lambda_{C\mid X}\left( u\mid X\right)\mid X,T\Big]\mid X\bigg]\Bigg]\Bigg|\\
    &=\Bigg|\mathbb{E}\Bigg[\int\mathbb{E}\bigg[ \frac{\mathbf{1}\{R(X,T)\ge\beta\} -\hat\eta(\beta,u\mid X)}{\hat{S}_{C\mid X}\left(u\mid X\right)}\mathbb{E}\Big[\mathbf{1}\{T\ge u\}\mathbf{1}\left\{C\ge u\right\}d\Lambda_{C\mid X}^\star\left( u\mid X\right)\\
    &\quad\quad  - \mathbf{1}\{T\ge u\}\mathbf{1}\{C\ge u\} d\hat\Lambda_{C\mid X}\left( u\mid X\right)\mid X,T\Big]\mid X\bigg]\Bigg]\Bigg|\\
    &=\Bigg|\mathbb{E}\Bigg[\int\mathbb{E}\Bigg[ \frac{\eta^\star(\beta,u\mid X)-\hat\eta(\beta,u\mid X)}{\hat{S}_{C\mid X}\left(u\mid X\right)}\mathbf{1}\{T\ge u\}S_{C\mid X}^\star(u\mid X)\\
    &\quad\quad d\left\{\Lambda_{C\mid X}^\star\left( u\mid X\right)-\hat\Lambda_{C\mid X}\left( u\mid X\right)\right\}\mid X\Bigg]\Bigg]\Bigg|\\
    &=\left|\mathbb{E}\left[\int \left\{\eta^\star(\beta,u\mid X)-\hat\eta(\beta,u\mid X)\right\}\frac{S_{C\mid X}^\star(u\mid X)}{\hat{S}_{C\mid X}\left(u\mid X\right)}S_{T\mid X}^\star(u\mid X)d\left\{\Lambda_{C\mid X}^\star\left( u\mid X\right)-\hat\Lambda_{C\mid X}\left( u\mid X\right)\right\}\right]\right|\\
    &=\left|\int\mathbb{E}\left[ \left\{\eta^\star(\beta,u\mid X)-\hat\eta(\beta,u\mid X)\right\}\frac{S_{C\mid X}^\star(u\mid X)}{\hat S_{C\mid X}\left(u\mid X\right)}S_{T\mid X}^\star(u\mid X)d\left\{\Lambda_{C\mid X}^\star\left( u\mid X\right)-\hat\Lambda_{C\mid X}\left( u\mid X\right)\right\}\right]\right|\\
    &< s_0 \int \left\| \hat\eta(\beta,u\mid X) - \eta^\star(\beta,u\mid X)\right\|_{L^2} \left\| d\left\{\hat\Lambda_{C\mid X}(u\mid X) - \Lambda_{C\mid X}^\star(u\mid X) \right\}\right\|_{L^2},
\end{align*}
where the last step follows from Cauchy-Schwarz inequality, and from bounding $S_{C\mid X}^\star$, $S_{T\mid X}^\star$ by $1$ and $\hat S_{C\mid X}^{-1}$ by $s_0$, using assumption \ref{Assum:bound}.
Moreover,
\begin{align*}
    &\left|\int\mathbb{E}\left[ \left\{\eta^\star(\beta,u\mid X)-\hat\eta(\beta,u\mid X)\right\}\frac{S_{C\mid X}^\star(u\mid X)}{\hat S_{C\mid X}\left(u\mid X\right)}S_{T\mid X}^\star(u\mid X)d\left\{\Lambda_{C\mid X}^\star\left( u\mid X\right)-\hat\Lambda_{C\mid X}\left( u\mid X\right)\right\}\right]\right|\\
    &=\left|\mathbb{E}\left[\int\left\{\eta^\star(\beta,u\mid X)-\hat\eta(\beta,u\mid X)\right\}\frac{S_{C\mid X}^\star(u\mid X)}{\hat S_{C\mid X}\left(u\mid X\right)}S_{T\mid X}^\star(u\mid X)d\left\{\Lambda_{C\mid X}^\star\left( u\mid X\right)-\hat\Lambda_{C\mid X}\left( u\mid X\right)\right\}\right]\right|\\
    &<s_0\left|\mathbb{E}\left[\int_0^{T\wedge C}\left\{\eta^\star(\beta,u\mid X)-\hat\eta(\beta,u\mid X)\right\}d\left\{\Lambda_{C\mid X}^\star\left( u\mid X\right)-\hat\Lambda_{C\mid X}\left( u\mid X\right)\right\}\right]\right|\\
    &=s_0\bigg|\mathbb{E}\left[\left\{\eta^\star(\beta,T\wedge C\mid X)-\hat\eta(\beta,T\wedge C\mid X)\right\}\left\{\Lambda_{C\mid X}^\star\left(T\wedge C\mid X\right)-\hat\Lambda_{C\mid X}\left(T\wedge C\mid X\right)\right\}\right]\\
    &\quad - \mathbb{E}\left[\int_0^{T\wedge C}\frac{d\left\{\eta^\star(\beta,u\mid X)-\hat\eta(\beta,u\mid X)\right\}}{du}\left\{\Lambda_{C\mid X}^\star\left( u\mid X\right)-\hat\Lambda_{C\mid X}\left( u\mid X\right)\right\}du\right]\bigg|\\
    &=s_0\bigg|\mathbb{E}\left[\left\{\eta^\star(\beta,Y\mid X)-\hat\eta(\beta,Y\mid X)\right\}\left\{\Lambda_{C\mid X}^\star\left(Y\mid X\right)-\hat\Lambda_{C\mid X}\left(Y\mid X\right)\right\}\right]\\
    &\quad - \int\mathbb{E}\left[d\left\{\eta^\star(\beta,u\mid X)-\hat\eta(\beta,u\mid X)\right\}S_{T\mid X}^\star(u\mid X)S_{C\mid X}^\star(u\mid X)\left\{\Lambda_{C\mid X}^\star\left( u\mid X\right)-\hat\Lambda_{C\mid X}\left( u\mid X\right)\right\}\right]\bigg|\\
    &\le s_0 \bigg\{\left\| \hat\eta(\beta,Y\mid X) - \eta^\star(\beta,Y\mid X)\right\|_{L^2} \left\|\hat\Lambda_{C\mid X}(Y\mid X) - \Lambda_{C\mid X}^\star(Y\mid X)\right\|_{L^2}\\
    &\quad + \int \left\| d\left\{\hat\eta(\beta,u\mid X) - \eta^\star(\beta,u\mid X)\right\}\right\|_{L^2} \left\| \hat\Lambda_{C\mid X}(u\mid X) - \Lambda_{C\mid X}^\star(u\mid X)\right\|_{L^2}\bigg\},
\end{align*}
where in the first inequality we bounded $\hat S_{C\mid X}^{-1}$ by $s_0$, using assumption \ref{Assum:bound}; and in the last step we used triangular inequality, Cauchy-Schwarz inequality, and we bounded $S_{C\mid X}^\star$, $S_{T\mid X}^\star$ by $1$. Therefore, for any $\beta\in[0,1]$,
\begin{align*}
    &\left|\mathbb{E}\left[\mathrm{IF}_\beta(\hat S_{C\mid X}, \hat\eta)\right]-\mathbb{E}\left[\mathrm{IF}_\beta(S_{C\mid X}^\star,\eta^\star)\right]\right|\\
    &<s_0 \min\bigg\{\int \left\| \hat\eta(\beta,u\mid X) - \eta^\star(\beta,u\mid X)\right\|_{L^2} \left\| d\left\{\hat\Lambda_{C\mid X}(u\mid X) - \Lambda_{C\mid X}^\star(u\mid X) \right\}\right\|_{L^2},\\
    &\quad \left\| \hat\eta(\beta,Y\mid X) - \eta^\star(\beta,Y\mid X)\right\|_{L^2} \left\|\hat\Lambda_{C\mid X}(Y\mid X) - \Lambda_{C\mid X}^\star(Y\mid X)\right\|_{L^2}\\
    &\quad + \int \left\| d\left\{\hat\eta(\beta,u\mid X) - \eta^\star(\beta,u\mid X)\right\}\right\|_{L^2} \left\| \hat\Lambda_{C\mid X}(u\mid X) - \Lambda_{C\mid X}^\star(u\mid X)\right\|_{L^2}\bigg\}.
\end{align*}
\end{proof}
Note that $\hat\Lambda_{C\mid X}$ and $\Lambda_{C\mid X}^\star$ are bounded above as a consequence of Assumption \ref{Assum:bound}.
%by $s_0\left(1+2\max\left\{1,\log s_0\right\}\right)$, as argued below in the proof of Lemma \ref{lem:bound-AIPCW}. 
Moreover, $\hat\eta$ and $\eta^\star$ are bounded as well, representing probabilities. Hence, the norms in equation (\ref{eq:DR}) are finite.

\section{Proof of Theorem \ref{thm:valid-AIPCW}}\label{App:5}
Similarly, to the proof of Theorem \ref{thm:valid-IPCW}, the proof of Theorem \ref{thm:valid-AIPCW} is based on two key results. First, Lemma \ref{lem:bound-AIPCW}, stated and proved below, which is similar to Lemma \ref{lem:bound-Op}. Second, Theorem \ref{thm:dbl_rob}. Recall the expression of the efficient influence function in equation (\ref{eq:IF}). Now define $\mathrm{IF}_\beta^i(S_{C\mid X},\eta)$ to be the influence function at the $i$-th observation $(X_i,Y_i,\Delta_i)$.

\begin{lemma}\label{lem:bound-AIPCW}
Under assumptions \ref{Assum:CIC} and \ref{Assum:bound}, for any $\epsilon>0$,
\begin{equation}\label{eq:bound-AIPCW}
\sup_{\beta\in[0, 1]}\,\left|\frac{1}{\left| \mathcal{D}_2\right| } \sum_{i \in \mathcal{I}_2} \mathrm{IF}_\beta^i(\hat S_{C\mid X},\hat\eta)- \mathbb{E}\left[\mathrm{IF}_\beta(\hat S_{C\mid X},\hat\eta)\mid  \mathcal{D}\right]\right|  \le \left(s_0+2\max\left\{1,s_0-1\right\}\right)\left(\left(\frac{1}{2}\log\frac{1}{\epsilon}\right)^{1/2}+K\right),
\end{equation}
with probability at least $1-\epsilon$, for some constant $K$.
\end{lemma}
\begin{proof}
Let $n=\left| \mathcal{D}_2\right| $. Let $f:\mathbb{R}^{d+2}\rightarrow\mathbb{R}$ be defined as
\[
f(x,\delta,t)=\frac{\delta\left\{\mathbf{1}\{R(x,t)\ge \beta\}-(1-\alpha)\right\}}{\hat{S}_{C\mid X}(t\mid x)} -\int \left\{\hat\eta(x,\beta,u) - (1-\alpha)\right\}\frac{d\hat M_{C\mid X}(u\mid x)}{\hat S_{C\mid X}(u\mid x)},
\]
given $\beta, R, \hat{S}_{C\mid X}, \hat\eta$ fixed. Notice that in this notation, we have explicitly included the dependence on $x$ in the functions $S_{C\mid X}$ and $\mu$, as highlighting this dependence is crucial for the subsequent steps. Like in the proof of Lemma \ref{lem:bound-Op}, we define the empirical process
\[
\mathbb G_n f := \frac{1}{\sqrt{n}} \sum_{i=1}^n\left(f(X_i,\Delta_i,T_i)-\mathbb{E}\left[f(X,\Delta,T)\mid\mathcal{D}\right]\right),
\]
and we want to show that for any $\epsilon> 0$ there exists $M(\epsilon)$ such that
\begin{equation*}
\mathbb P\left(\sup_{\beta\in[0, 1]}\left|\mathbb G_n f \right| \ge M(\epsilon)\right)\le \epsilon.
\end{equation*}
To apply McDiarmid's inequality to the function $\sup_\beta\left| \mathbb G_n f \right|$, we verify the bounded differences property. Given $j\in[n]$ fixed,
\begin{align*}
    &\left| f(x_j,\delta_j,t_j) - f(x_j',\delta_j',t_j') \right| \\
    &\quad = \bigg|\frac{\delta_j\left\{\mathbf{1}\{R(x_j,t_j)\ge \beta\}-(1-\alpha)\right\}}{\hat{S}_{C\mid X}(t_j\mid x_j)} -\int \left\{\hat\eta(x_j,\beta,u) -(1-\alpha)\right\}\frac{d\hat M_{C\mid X}(u\mid x_j)}{\hat S_{C\mid X}(u\mid x_j)}\\
    &\quad\quad - \frac{\delta_j'\left\{\mathbf{1}\{R(x_j',t_j')\ge \beta\}-(1-\alpha)\right\}}{\hat{S}_{C\mid X}(t_j'\mid x_j')} +\int \left\{\hat\eta(x_j',\beta,u) -(1-\alpha)\right\}\frac{d\hat M_{C\mid X}(u\mid x_j')}{\hat S_{C\mid X}(u\mid x_j')}\bigg|\\
    &\quad < s_0 + \left|\int \left\{\hat\eta(x_j,\beta,u)-(1-\alpha) \right\}\frac{d\hat M_{C\mid X}(u\mid x_j)}{\hat S_{C\mid X}(u\mid x_j)}\right|+ \left|\int \left\{\hat\eta(x_j',\beta,u)-(1-\alpha) \right\}\frac{d\hat M_{C\mid X}(u\mid x_j')}{\hat S_{C\mid X}(u\mid x_j')}\right|,
\end{align*}
where we applied the triangular inequality and the bound on the IPCW term of $f$ recovered in the proof of Lemma \ref{lem:bound-Op}. Now we bound the second and third terms in the expression above:
\begin{align*}
    \left|\int \left\{\hat\eta(x,\beta,u)-(1-\alpha) \right\}\frac{d\hat M_{C\mid X}(u\mid x)}{\hat S_{C\mid X}(u\mid x)}\right|
    &\le\int \frac{\left|d\hat M_{C\mid X}(u\mid x)\right|}{\hat S_{C\mid X}(u\mid x)}\le \max\left\{1,s_0-1\right\},
\end{align*}
by equation \ref{eq:M} and Assumption \ref{Assum:bound}.
%since, by assumption \ref{Assum:bound}, $\hat\Lambda_{C\mid X}(u\mid X)\in\left[0, \log s_0\right]$.
Therefore, 
\begin{align*}
    &\sup_{\substack{(x_1,\delta_1,t_1),\dots, (x_n,\delta_n,t_n) \\  (x_j',\delta_j',t_j')}}\Bigg|\sup_{\beta\in[0, 1]}\Bigg|\frac{1}{\sqrt{n}} \sum_{i=1}^n\left(f(x_i,\delta_i,t_i) - \mathbb{E}\left[f(X, \Delta,T)\mid\mathcal{D}\right]\right)\Bigg|\\
    &\quad -\sup_{\beta\in[0, 1]}\Bigg|\frac{1}{\sqrt{n}} \sum_{i\neq j} \left(f(x_i,\delta_i,t_i) - \mathbb{E}\left[f(X,\Delta,T)\mid\mathcal{D}\right]\right)+\frac{1}{\sqrt{n}} \left(f(x_j',\delta_j',t_j') - \mathbb{E}\left[f(X,\Delta,T)\mid\mathcal{D}\right]\right)\Bigg|\Bigg|\\
    &\quad \leq \sup_{\substack{(x_1,\delta_1,t_1),\dots, (x_n,\delta_n,t_n) \\  (x_j',\delta_j',t_j')}}\sup_{\beta\in[0, 1]} \frac{1}{\sqrt{n}}\left| f(x_j,\delta_j,t_j) - f(x_j',\delta_j',t_j') \right|\\
    &\quad < \frac{s_0 + 2\max\left\{1, s_0-1\right\}}{\sqrt{n}}.
\end{align*}
Hence, we can apply McDiarmid's inequality which yields
\begin{align*}
    &\mathbb P\left(\sup_{\beta\in[0, 1]}\left| \mathbb G_n f \right|  - \mathbb{E}\left[\sup_{\beta\in[0, 1]}\left| \mathbb G_n f \right|\right] \ge v\right)\le \exp\left(-\frac{2v^2}{\sum_{i=1}^n \left\{\left(s_0+2\max\left\{1,s_0-1\right\}\right)\right\}^2/n}\right)\\
    &=\exp\left(-\frac{2v^2}{\left(s_0+2\max\left\{1,s_0-1\right\}\right)^2}\right),
\end{align*}
for any $v>0$. Following the same steps as in the proof of Lemma \ref{lem:bound-Op}, we obtain the result.
\end{proof}

\begin{proof}[Proof of Theorem \ref{thm:valid-AIPCW}]
Fix $\epsilon\in(0,1)$. For any fixed $\beta\in[0,1]$,
\begin{align*}
    &\left|\frac{1}{\left| \mathcal{D}_2\right| }\sum_{i\in\mathcal{I}_2} \mathrm{IF}_\beta^i(\hat S_{C\mid X},\hat\eta)- \left\{\mathbb{P}\left(R(X,T)\ge\beta\mid \mathcal{D}\right)-(1-\alpha)\right\} \right|\\
    &\overset{(\ref{eq:IPCW})}=\left|\frac{1}{\left| \mathcal{D}_2\right| }\sum_{i\in\mathcal{I}_2} \mathrm{IF}_\beta^i(\hat S_{C\mid X},\hat\eta) -\mathbb{E}\left[\frac{\Delta\mathbf{1}\{R(X,T)\ge\beta\}}{S_{C\mid X}^\star(T\mid X)}\mid \mathcal{D}\right]+(1-\alpha)\right|\\
    &=\left|\frac{1}{\left| \mathcal{D}_2\right| }\sum_{i\in\mathcal{I}_2} \mathrm{IF}_\beta^i(\hat S_{C\mid X},\hat\eta) -\mathbb{E}\left[\frac{\Delta\mathbf{1}\{R(X,T)\ge\beta\}}{S_{C\mid X}^\star(T\mid X)}\mid \mathcal{D}\right]+(1-\alpha)\mathbb{E}\left[\frac{\Delta}{S_{C\mid X}^\star(T\mid X)}\mid \mathcal{D}\right]\right|\\
    &=\left|\frac{1}{\left| \mathcal{D}_2\right| }\sum_{i\in\mathcal{I}_2}\mathrm{IF}_\beta^i(\hat S_{C\mid X},\hat\eta)-\mathbb{E}\left[\frac{\Delta\left\{\mathbf{1}\{R(X,T)\ge\beta\}-(1-\alpha)\right\}}{S_{C\mid X}^\star(T\mid X)}\mid \mathcal{D}\right]\right|\\
    &\le\sup_{\gamma\in[0, 1]}\Bigg|\frac{1}{\left| \mathcal{D}_2\right| }\sum_{i\in\mathcal{I}_2}\mathrm{IF}_\gamma^i(\hat S_{C\mid X},\hat\eta)-\mathbb{E}\left[\frac{\Delta\left\{\mathbf{1}\{R(X,T)\ge\gamma\}-(1-\alpha)\right\}}{S_{C\mid X}^\star(T\mid X)}\mid \mathcal{D}\right]\\
    &\quad -\mathbb{E}\left[\int \frac{d M_{C\mid X}^\star(u\mid x)}{S_{C\mid X}^\star(u\mid x)}\left\{\eta^\star(\gamma,u\mid X) -(1-\alpha) \right\}du\right]\Bigg|\\
    &=\sup_{\gamma\in[0, 1]}\left|\frac{1}{\left| \mathcal{D}_2\right| }\sum_{i\in\mathcal{I}_2}\mathrm{IF}_\gamma^i(\hat S_{C\mid X},\hat\eta)- \mathbb{E}\left[\mathrm{IF}_\gamma(S_{C\mid X}^\star,\eta^\star)\mid  \mathcal{D}\right]\right|\\
    &\le \sup_{\gamma\in[0, 1]}\,\left|\frac{1}{\left| \mathcal{D}_2\right| }\sum_{i\in\mathcal{I}_2}\mathrm{IF}_\gamma^i(\hat S_{C\mid X},\hat\eta)- \mathbb{E}\left[\mathrm{IF}_\gamma(\hat S_{C\mid X},\hat\eta)\mid  \mathcal{D}\right]\right|\\
    &\quad +\sup_{\gamma\in[0, 1]}\left| \mathbb{E}\left[\mathrm{IF}_\gamma(\hat S_{C\mid X},\hat\eta)\mid  \mathcal{D}\right]- \mathbb{E}\left[\mathrm{IF}_\gamma(S_{C\mid X}^\star,\eta^\star)\mid  \mathcal{D}\right]\right|\\
    &\stackrel{\mathrm{(\ref{eq:bound-AIPCW}),(\ref{eq:DR})}}< \left(s_0+2\max\left\{1,s_0-1\right\}\right)\left(\left(\frac{1}{2}\log\frac{1}{\epsilon}\right)^{1/2}+K\right)\frac{1}{\sqrt{\left| \mathcal{D}_2\right| }}\\
    &\quad\quad + s_0 \min\Bigg\{\int \sup_{\beta\in[0,1]} \left\| \hat\eta(\beta,u\mid X) - \eta^\star(\beta,u\mid X)\right\|_{L^2} \left\| d\left\{\hat\Lambda_{C\mid X}(u\mid X) - \Lambda_{C\mid X}^\star(u\mid X) \right\}\right\|_{L^2},\\
    &\quad\quad \left\| \hat\eta(\beta,Y\mid X) - \eta^\star(\beta,Y\mid X)\right\|_{L^2} \left\|\hat\Lambda_{C\mid X}(Y\mid X) - \Lambda_{C\mid X}^\star(Y\mid X)\right\|_{L^2}\\
    &\quad\quad + \int \sup_{\beta\in[0,1]} \left\| d\left\{\hat\eta(\beta,u\mid X) - \eta^\star(\beta,u\mid X)\right\}\right\|_{L^2} \left\| \hat\Lambda_{C\mid X}(u\mid X) - \Lambda_{C\mid X}^\star(u\mid X)\right\|_{L^2}\Bigg\}
\end{align*}
with probability at least $1-\epsilon$. The result follows by letting $\beta=\hat{\beta}_{\mathrm{AIPCW}}$ in the inequality above, and because $\hat{\beta}_{\mathrm{AIPCW}}$ satisfies $\frac{1}{\left| \mathcal{D}_2\right| }\sum_{i\in\mathcal{I}_2}\mathrm{IF}_{\hat{\beta}_{\mathrm{AIPCW}}}^i(\hat S_{C\mid X},\hat\eta) \ge 0$ by construction.
\end{proof}

\newpage
\section{Algorithms}\label{App:algo}
\begin{algorithm}[!h]
\caption{IPCW method to estimate the LPB based on the conditional quantile}
\label{algo:IPCW}
\KwIn{Dataset $\mathcal{D}$, level $\alpha$, grid of points $\left\{\beta_j\right\}_{j\in\mathcal{J}}$ in $[0,1]$}
Partition the data $\mathcal{D}$ into a training set $\mathcal{D}_1$ and a calibration set $\mathcal{D}_2$, with respective index sets $\mathcal{I}_1$ and $\mathcal{I}_2$\;
On the training set $\mathcal{D}_1$, fit the survival functions $\hat{S}_{C\mid X}(\cdot\mid  \cdot)$ and $\hat{S}_{T\mid X}(\cdot\mid  \cdot)$ using any appropriate algorithm(s)\;
Based on the estimated survival curve $\hat{S}_{T\mid X}(t\mid X)$, obtain the conditional $\beta$-quantile $\hat{q}_{T\mid X}(\beta_j\mid X_i)$ for all $j\in\mathcal{J}$ and for all $i\in\mathcal{I}_2$\;
\For{$j\in\mathcal{J}$}{
    Compute the estimated coverage probability
    \[
    \hat{W}(\beta_j) = \frac{1}{\left| \mathcal{D}_2\right| } \sum_{i \in \mathcal{I}_2} 
    \frac{\Delta_i \left\{\mathbf{1}\{T_i \ge \hat{q}_{T\mid X}(\beta_j\mid X_i)\} - (1-\alpha)\right\}}{\hat{S}_{C\mid X}(T_i \mid X_i)}
    \]
}
Compute $\hat{\beta}_{\mathrm{IPCW}} = \sup \left\{ \beta_j, j\in\mathcal{J} : \hat{W}(\beta_j) \geq 0 \right\}$\;
\KwOut{$\hat{L}(\cdot) = \hat{q}_{T\mid X}(\hat{\beta}_{\mathrm{IPCW}}\mid \cdot)$}
\end{algorithm}

\begin{algorithm}[!h]
\caption{IPCW method to estimate the LPB based on a general non-conformity score}
\label{algo:IPCW-R}
\KwIn{Dataset $\mathcal{D}$, level $\alpha$, grid of points $\left\{\beta_j\right\}_{j\in\mathcal{J}}$ in $[0,1]$, non-conformity score $R(\cdot,\cdot)$}
Partition the data $\mathcal{D}$ into a training set $\mathcal{D}_1$ and a calibration set $\mathcal{D}_2$, with respective index sets $\mathcal{I}_1$ and $\mathcal{I}_2$\;
On the training set $\mathcal{D}_1$, fit the survival functions $\hat{S}_{C\mid X}(\cdot\mid  \cdot)$ and $R(\cdot,\cdot)$ using any appropriate algorithm(s)\;
Based on the non-conformity score $R(\cdot,\cdot)$, compute the score $R( X_i, T_i)$ for all $i\in\mathcal{I}_2$ for which $\Delta_i=1$\;
\For{$j\in\mathcal{J}$}{
    Compute the estimated coverage probability 
    \[
    \hat{W}(\beta_j) = \frac{1}{\left| \mathcal{D}_2\right| } \sum_{i \in \mathcal{I}_2} \frac{\Delta_i \left\{\mathbf{1}\{R(X_i,T_i) \ge \beta_j\} - (1-\alpha)\right\}}{\hat{S}_{C\mid X}(T_i \mid  X_i)}
    \]
}
Compute $\hat{\beta}_{\mathrm{IPCW}} = \sup \left\{ \beta_j, j\in\mathcal{J} : \hat{W}(\beta_j) \geq 0 \right\}$\;
\KwOut{Prediction region given a new $X$: $\left\{t:R(X,t)\ge\hat{\beta}_{\mathrm{IPCW}}\right\}$}
\end{algorithm}

\begin{algorithm}
\caption{AIPCW method to estimate the LPB based on the conditional quantile}
\label{algo:AIPCW}
\KwIn{Dataset $\mathcal{D}$, level $\alpha$, grid of points $\left\{\beta_j\right\}_{j\in\mathcal{J}}$ in $[0,1]$}
Partition the dataset $\mathcal{D}$ into a training set $\mathcal{D}_1$ and a calibration set $\mathcal{D}_2$, with respective index sets $\mathcal{I}_1$ and $\mathcal{I}_2$\;
Fit the survival functions $\hat{S}_{C\mid X}(\cdot\mid  \cdot)$ and $\hat{S}_{T\mid X}(\cdot\mid  \cdot)$ on $\mathcal{D}_1$ using an appropriate algorithm\;
Obtain the conditional $\beta$-quantile $\hat{q}_{T\mid X}(\beta_j\mid X_i)$ from $\hat{S}_{T\mid X}(t\mid X)$ for all $j\in\mathcal{J}$ and $i\in\mathcal{I}_2$\;
Let $u_{(1)},\dots,u_{(Q)}$ be the observed ordered censoring times, with $u_{(0)}=0$. Based on the estimated survival curve $\hat{S}_{T\mid X}(t\mid X)$, obtain
\[
\hat\eta(\beta_j,u_{(k)}\mid X_i) = \frac{\hat S_{T\mid X}\left( \max\{\hat q_{T\mid X}(\beta_j\mid X_i), u_{(k)}\} \mid X_i \right)}{\hat S_{T\mid X}\left( u_{(k)} \mid X_i\right)}
\]
for all $j\in\mathcal{J}$, for all $k=0,\dots,Q$, and for all $i\in\mathcal{I}_2$\;
Based on the estimated survival curve $\hat{S}_{C\mid X}(t\mid X)$, obtain
\[
\hat M_{C\mid X}\left(u_{(k)}\mid X_i\right)=\mathbf{1}\{Y_i\le u_{(k)},\Delta_i =0\} + \log \hat{S}_{C\mid X}\left(Y\land u_{(k)}\mid X_i\right) 
\]
for all $j\in\mathcal{J}$, for all $k=0,\dots,Q$, and for all $i\in\mathcal{I}_2$\;
\For{$j\in\mathcal{J}$}{
    Compute the estimated IPCW and AIPCW terms
    \[
    \hat{W}(\beta_j) = \frac{1}{\left| \mathcal{D}_2\right| } \sum_{i \in \mathcal{I}_2} \frac{\Delta_i \left\{\mathbf{1}\{T_i \ge \hat{q}_{T\mid X}(\beta_j\mid X_i)\} - (1-\alpha)\right\}}{\hat{S}_{C\mid X}(T_i \mid  X_i)}
    \]
    \begin{align*}
        \hat{\Pi}(\beta_j)&=\frac{1}{\left| \mathcal{D}_2\right| }\sum_{i\in \mathcal{I}_2 } \sum_{k=1}^Q \frac{\hat\eta(\beta_j,u_{(k)}\mid X_i)-(1-\alpha)}{\hat{S}_{C\mid X}(u_{(k)}\mid X_i)}\\
        &\quad\cdot\left\{\hat M_{C\mid X}\left(u_{(k)}\mid X_i\right)-\hat M_{C\mid X}\left(u_{(k-1)}\mid X_i\right)\right\}
    \end{align*}
}
Compute $\hat{\beta}_{\mathrm{AIPCW}} = \sup \left\{ \beta_j, j\in\mathcal{J} : \hat{W}(\beta_j) + \hat{\Pi}(\beta_j)\ge 0 \right\}$\;
\KwOut{$\hat{L}(\cdot) = \hat{q}_{T\mid X}(\hat{\beta}_{\mathrm{AIPCW}}\mid \cdot)$}
\end{algorithm}

\begin{algorithm}
\caption{AIPCW method to estimate the LPB based on a general non-conformity score}
\label{algo:AIPCW-R}
\KwIn{Dataset $\mathcal{D}$, level $\alpha$, grid of points $\left\{\beta_j\right\}_{j\in\mathcal{J}}$ in $[0,1]$, non-conformity score $R(\cdot,\cdot)$}
Partition the dataset $\mathcal{D}$ into a training set $\mathcal{D}_1$ and a calibration set $\mathcal{D}_2$, with respective index sets $\mathcal{I}_1$ and $\mathcal{I}_2$\;
On the training set $\mathcal{D}_1$, fit the survival functions $\hat{S}_{C\mid X}(\cdot\mid  \cdot)$ and $R(\cdot,\cdot)$ using any appropriate algorithm(s)\;
Based on the non-conformity score $R(\cdot,\cdot)$, compute the score $R( X_i, T_i)$ for all $i\in\mathcal{I}_2$ for which $\Delta_i=1$\;
Let $u_{(1)},\dots,u_{(Q)}$ be the observed ordered censoring times, and $u_{(0)}=0$. Based on the estimated non-conformity score $R(\cdot,\cdot)$, obtain the estimate $\hat\eta(\beta_j,u_{(k)}\mid X_i)$, for all $j\in\mathcal{J}$, for all $k=0,\dots,Q$, and for all $i\in\mathcal{I}_2$\;
Based on the estimated survival curve $\hat{S}_{C\mid X}(t\mid X)$, obtain
\[
\hat M_{C\mid X}\left(u_{(k)}\mid X_i\right)=\mathbf{1}\{ Y_i\le u_{(k)},\Delta_i =0\} + \log \hat{S}_{C\mid X}\left(Y\land u_{(k)}\mid X_i\right) 
\]
for all $j\in\mathcal{J}$, for all $k=0,\dots,Q$, and for all $i\in\mathcal{I}_2$\;
\For{$j\in\mathcal{J}$}{
    Compute the estimated IPCW and AIPCW terms 
    \[
    \hat{W}(\beta_j) = \frac{1}{\left| \mathcal{D}_2\right| } \sum_{i \in \mathcal{I}_2} \frac{\Delta_i \left\{\mathbf{1}\{R(X_i,T_i) \ge \beta_j\} - (1-\alpha)\right\}}{\hat{S}_{C\mid X}(T_i \mid  X_i)}
    \]
    \begin{align*}
        \hat{\Pi}(\beta_j)&=\frac{1}{\left| \mathcal{D}_2\right| }\sum_{i\in \left| \mathcal{D}_2\right| } \sum_{k=1}^Q\frac{\hat\eta(\beta_j,u_{(k)}\mid X_i)-(1-\alpha)}{\hat{S}_{C\mid X}(u_{(k)}\mid X_i)}\\
        &\quad \cdot\left\{\hat M_{C\mid X}\left(u_{(k)}\mid X_i\right)-\hat M_{C\mid X}\left(u_{(k-1)}\mid X_i\right)\right\}
    \end{align*}
}
$\hat{\beta}_{\mathrm{AIPCW}} = \sup \left\{ \beta_j, j\in\mathcal{J} : \hat{W}(\beta_j) + \hat{\Pi}(\beta_j)\ge 0 \right\}$\;
\KwOut{Prediction region given a new $X$: $\left\{t:R(X,t)\ge\hat{\beta}_{\mathrm{AIPCW}}\right\}$}
\end{algorithm}

\newpage
\section{Competing Methods and Implementation Details}\label{App:comp}
This section describes the additional benchmark and competing methods included in our simulation study, and summarizes how we implemented them so that all approaches are evaluated under the same data splits and metrics. All methods are trained on the training split, calibrated (when applicable) on the calibration split, and evaluated on the test split using empirical coverage and the average LPB.

As reference points, we consider simple quantile-based baselines that ignore the censoring mechanism. These naive benchmark methods are:
\begin{itemize}
    \item Quantile Regression forests on $Y$ (QR-Y): Conditional quantile regression forests at level $\alpha$ applied to $(X_i, Y_i)$, returning the predicted quantile as the LPB. Since any LPB on $Y = \min\{T, C\}$ is also a LPB on $T$, this method does not provide any coverage guarantee as it fails to appropriately account for censoring as a nuisance parameter.
    \item Conformalized quantile regression forests on $Y$ (CQR-Y): Split-CQR \citep{Romano} applied to $(X_i, Y_i)$ using quantile regression forests. Although this approach produces overly-conservative bounds, it can be useful in settings where the censoring mechanism is unknown or the conditionally independent censoring assumption does not hold, provided the quantile regression model is sufficiently flexible to be consistent. 
    \item Quantile Regression forests on $T$ (QR-T): Conditional quantile regression forests at level $\alpha$ applied to $(X_i, \Delta_iT_i)$, using the predicted quantile as the LPB. Because it only considers uncensored data points, it may be overly conservative as it discards information contained in censored observations and completely relies on a consistent estimator of the conditional quantile function. 
    \item Conformalized quantile regression forests on $T$ (CQR-T): Split-CQR \citep{Romano} applied to $(X_i, \Delta_iT_i)$ using quantile regression forests. Similar to QR-T, this method also fails to leverage information contained in censored observations.
\end{itemize}
All these are implemented via the \texttt{quantregForest} package \citep{meinshausen2017package}.

As competing methods, we also consider the following recent conformal survival methods:
\begin{itemize}
    \item \citet{sesia2025doubly} propose a conformal method for standard right-censoring that constructs LPBs for $T$ by combining imputation of latent censoring times with weighted conformal calibration. Their method first imputes censoring times for individuals with observed time-to-event under conditionally independent censoring to create a pseudo Type-I censoring dataset, and then applies split conformal inference with inverse censoring weights (with an adaptive-cutoff option). In our implementation, we used the authors' code available at \url{https://github.com/msesia/conformal_survival}. We fit both the event and censoring nuisance models using their default learner (RSF), via the \texttt{R} package \texttt{randomForestSRC} \citep{ishwaran2021fast}. 
    \item \citet{qin2025conformal} propose conformal predictive intervals for survival analysis under standard right-censoring using a bootstrap calibration strategy. Their framework supports one-sided and two-sided predictive intervals for $T$, calibrated through bootstrap resampling. In our experiments, we used their public code at \url{https://github.com/JPiao7u089/CPI-bootstrap}. The reference implementation fits a Cox proportional hazards model for the event time and uses a Kaplan-Meier estimator for the censoring distribution, and then applies a bootstrap procedure to calibrate conformal thresholds on the survival scale. The one-sided LPB is obtained by mapping the calibrated survival cutoff back to time via inversion of the fitted Cox baseline cumulative hazard.
    \item \citet{si2025training} propose a procedure for constructing LPBs under standard right-censoring with a training-set conditional APAC-type validity target, i.e.,
    \[
    \mathbb{P}\left(\mathbb{P}(T \ge \hat{L}(X)\mid\mathcal{D}) \ge 1 - \alpha\right) \ge 1 - \epsilon - o(1) \quad \mbox{as}\quad n\to\infty.
    \]
    The procedure proposed by \citet{si2025training} follows a two-stage template closely related to our AIPCW construction: it first learns the event-time and censoring mechanisms, and then performs a one-step (efficient-influence-function) correction to calibrate the prediction bound. The calibration step differs from ours: their LPB is selected through a Wald lower confidence bound aimed at APAC validity. In practice, the final LPB is obtained by evaluating an estimated survival curve for \(T\mid X\) and inverting it at the calibrated level. We used the authors' implementation available at \url{https://github.com/averysi224/tc-surv}. In our experiments, we fit the event and censoring nuisance components with \texttt{survSuperLearner}. In terms of the assumptions required for validity, Assumption A4 of~\citet{si2025training} implies that the APAC guarantee does not hold unless an appropriate product of nuisance-estimation errors converges to zero faster than \(n^{-1/2}\), which can be a restrictive condition, particularly in high-dimensional settings. By contrast, for our AIPCW-based method, Theorem~\ref{thm:valid-AIPCW} shows that the \(o_p(1)\) remainder term in our coverage guarantee (see Definition~\ref{def:PAC}) vanishes whenever the same product of errors converges to zero in probability, without requiring any specific rate of convergence.
    \item \citet{yi2025survival} develop survival conformal prediction under random right-censoring using inverse censoring weighting within a weighted conformal calibration scheme. Their key idea is to correct the non-exchangeability induced by censoring via IPCW, combined with a nonconformity score based on censored quantile regression, yielding prediction intervals for $T$. We used the original implementation kindly provided by the authors (personal communication). Following their code, we estimated censoring weights from the training data using kernel-based routines with data-driven bandwidths, capped extreme inverse-weight values for stability, computed weighted conformal $p$-values over a grid of candidate times, and took the lower endpoint of the resulting interval (equivalently, the minimum accepted time on the grid) as the reported LPB. For numerically sensitive settings, we also applied grid truncation and a positivity floor for inverse weights, as suggested by the authors.
    \item \citet{meixide} propose a distribution-free uncertainty-quantification method for interval-censored outcomes. The method returns prediction sets for interval-censored targets and provides two operating modes: a conservative finite-sample mode and an asymptotic randomized mode (denoted $e=0$ and $e=*$ in their notation). Since the method is built for interval censoring rather than standard right-censoring, comparison in our setting is necessarily an adaptation. We followed a best-effort adaptation by mapping right-censoring to interval-censoring via $(L,U)=(Y,\infty)$. We use the authors' implementation available at \url{https://github.com/meixide/uncervals}, and run our simulation in mode $e=*$. For the required base interval-censored survival estimator, we used Interval Censored Recursive Forest, as suggested by the paper.
\end{itemize}

\section{Simulation settings}\label{App:simu_set}
The settings of our synthetic experiments are outlined in Table \ref{tab:simu_setting}.
\begin{table}[h]
    \centering
    \caption{Summary of the settings for the synthetic experiments.}
    \begin{tabular}{ccl}
        \toprule
        Setting & $p$ & $P_X$, $P_{T\mid X}$,$P_{C\mid X}$ \\ \midrule
        1 & 2 &
        \begin{tabular}[t]{@{}l@{}}
            $X\sim N(0,1)^{2}$\\
            $T\mid X\sim \mathrm{Exp}\left(\exp(-X_1+X_2)\right)$\\
            $C\mid X\sim \mathrm{Exp}\!\left(1/3\right)$
        \end{tabular} \\ \midrule
        2 & 10 &
        \begin{tabular}[t]{@{}l@{}}
            $X\sim N(0,1)^{10}$\\
            $T\mid X\sim \mathrm{Exp}\left(2\exp\left((X_1X_2 - X_3^2)/3\right)/3\right)$\\
            $C\mid X\sim \mathrm{Exp}\!\left(2\exp\left((X_3 - X_4^2)/3\right)/3\right)$
        \end{tabular} \\ \midrule
        3 & 100 &
        \begin{tabular}[t]{@{}l@{}}
            $X\sim \mathrm{Unif}[-1,1]^{100}$\\
            $T\mid X\sim \mathrm{LogNorm}(\mathbf{1}\{X\in A_T\}\log(10) + \mathbf{1}\{X\notin A_T\}\log(1000), 1)$\\
            $A_T=\left\{X_i>0,\, X_j<0 \ \ i=1,\dots,5,\, j=6,\dots,10\right\}$\\
            $C\mid X\sim \mathrm{LogNorm}(\mathbf{1}\{X\in A_C\}\log(10) + \mathbf{1}\{X\notin A_C\}\log(1000), 1)$\\
            $A_C=\left\{X_1>0,\, X_2<0\right\}$
        \end{tabular} \\ \midrule
        4 & 100 &
        \begin{tabular}[t]{@{}l@{}}
            $X\sim \mathrm{Unif}[-1,1]^{100}$\\
            $T\mid X\sim \mathrm{LogNorm}(\mathbf{1}\{X\in A\}\log(10) + \mathbf{1}\{X\notin A\}\log(1000), 1)$\\
            $A=\left\{X_2<0, X_3>0, X_4>0\right\}$\\
            $C\mid X\sim \mathrm{LogNorm}(\mathbf{1}\{X_1<0\}\log(10) + \mathbf{1}\{X_1\ge 0\}\log(1000), 1)$
        \end{tabular} \\ \midrule
        5 & 100 &
        \begin{tabular}[t]{@{}l@{}}
            $X\sim \mathrm{Unif}[0,1]^{100}$\\
            $T\mid X\sim \mathrm{LogNorm}\big((X_1-0.5)^2 + X_2X_3 - \mathbf{1}\{X_3<0.5,\, X_4>0.5\} + $\\
            $\quad\quad\quad\quad (X_5)^{1/2} + (X_6+X_7-0.5)^3,\, 1\big)$\\
            $C\mid X\sim \mathrm{LogNorm}\big((X_1+X_2-1)^2 - X_3X_4 + \mathbf{1}\{X_6>0.5\} -$\\
            $\quad\quad\quad\quad(X_7-0.5)^3 X_8,\, 1\big)$
        \end{tabular} \\ \midrule
        6 & 100 &
        \begin{tabular}[t]{@{}l@{}}
            $X\sim \mathrm{Unif}[0,1]^{100}$\\
            $T\mid X\sim \mathrm{LogNorm}\!\left(0.126\left(X_1 + (X_3X_5)^{1/2}\right)+1,\; (X_2+2)/4\right)$\\
            $C\mid X\sim \mathrm{Exp}\!\left(X_6/2\right)$
        \end{tabular} \\ \bottomrule
    \end{tabular} 
    \label{tab:simu_setting}
\end{table}

\section{Additional simulation results}\label{App:plots}

\begin{figure}[h!]
    \centering
    \includegraphics[width=\textwidth]{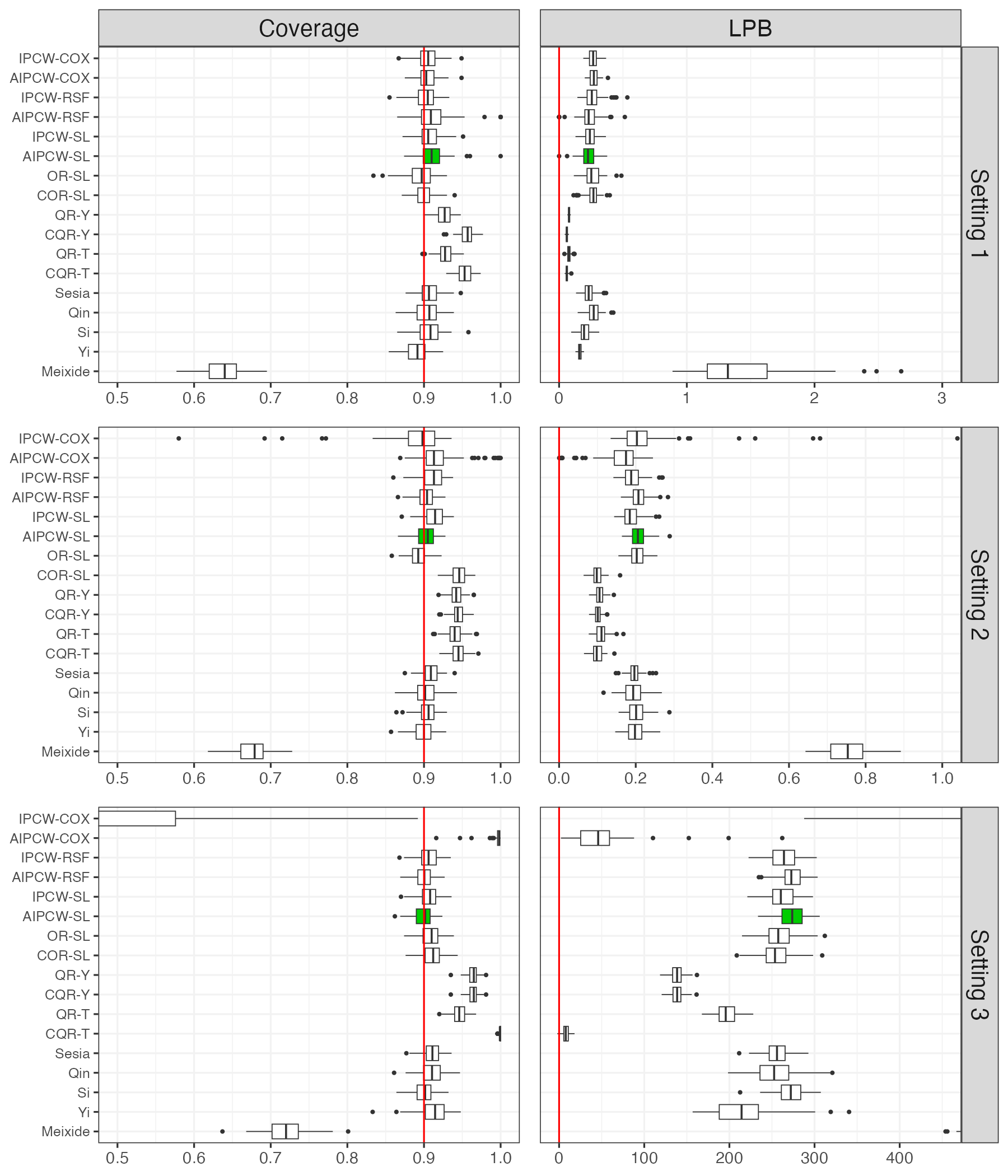}
    \caption{Distribution of empirical marginal coverage rates (left panel) and average estimated LPB (right panel) across 100 simulated test datasets under Settings 1,2,3 for each method. The red vertical line in the left panel denotes the 90\% target coverage, while the red vertical line in the right panel marks the lower limit of the support of the time-to-event outcome. The recommended method AIPCW-SL is highlighted in green.}
    \label{fig:coverage_boxplot_setting1to3}
\end{figure}

\begin{figure}[h!]
    \centering
    \includegraphics[width=\textwidth]{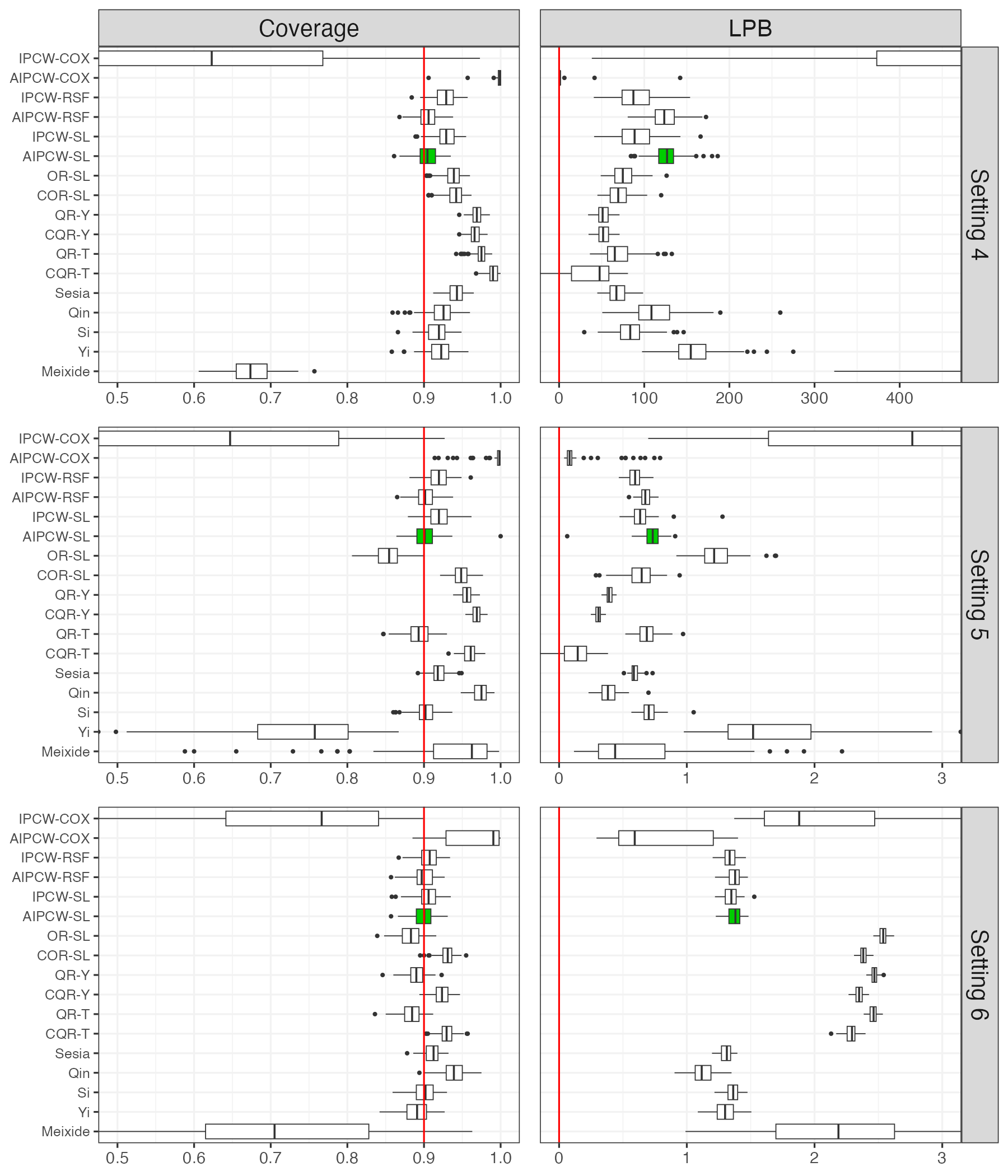}
    \caption{Distribution of empirical marginal coverage rates (left panel) and average estimated LPB (right panel) across 100 simulated test datasets under Settings 4,5,6 for each method. The red vertical line in the left panel denotes the 90\% target coverage, while the red vertical line in the right panel marks the lower limit of the support of the time-to-event outcome. The recommended method AIPCW-SL is highlighted in green.}
    \label{fig:coverage_boxplot_setting4to6}
\end{figure}

Figures~\ref{fig:coverage_boxplot_setting1to3} and~\ref{fig:coverage_boxplot_setting4to6} provide a closer examination of the empirical coverage rates and average LPBs for each of the six experimental settings (see Figures~\ref{fig:coverage_boxplot} and~\ref{fig:lpb_boxplot}). With respect to the main text, the following visualizations facilitate a deeper understanding of the performance of each method under different scenarios, highlighting their strengths and limitations in terms of coverage accuracy. We refer to section \ref{sec:simu} for description and interpretation of these plots.

\begin{figure}[t]
    \centering
    \includegraphics[width=\textwidth]{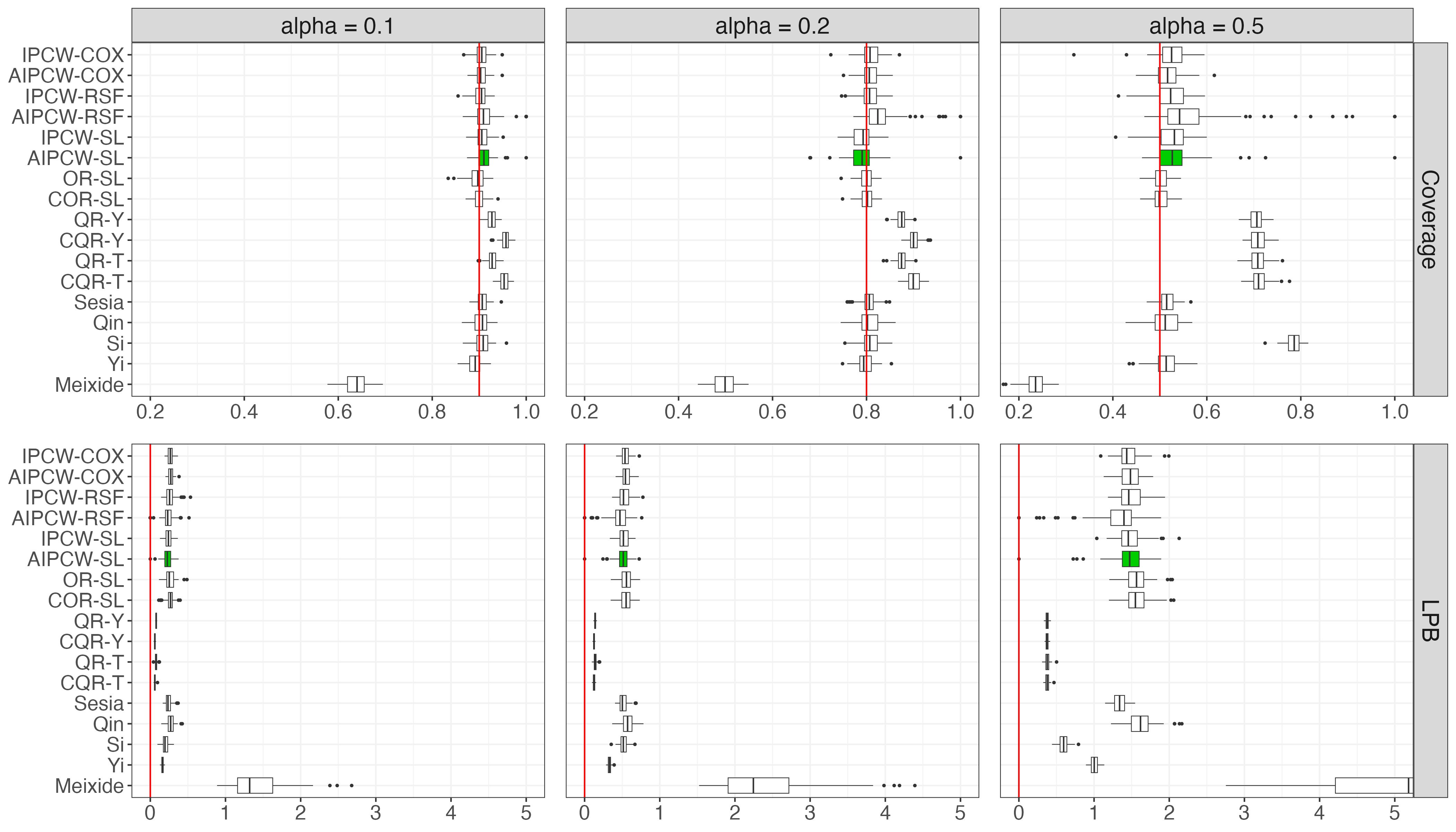}
    \caption{Distribution of empirical marginal coverage rates (top row) and average estimated LPB (bottom row) across 100 simulated test datasets under Setting 1 for each method, with different miscoverage levels $\alpha$. The red vertical lines in the top row denote the nominal coverage levels of 90\%, 80\%, and 50\% (from left to right), while the red vertical lines in the bottom row panel marks the lower limit of the support of the time-to-event outcome. The recommended method AIPCW-SL is highlighted in green.}
    \label{fig:boxplot_alpha}
\end{figure}

\begin{figure}[h!]
    \centering
    \includegraphics[width=\textwidth]{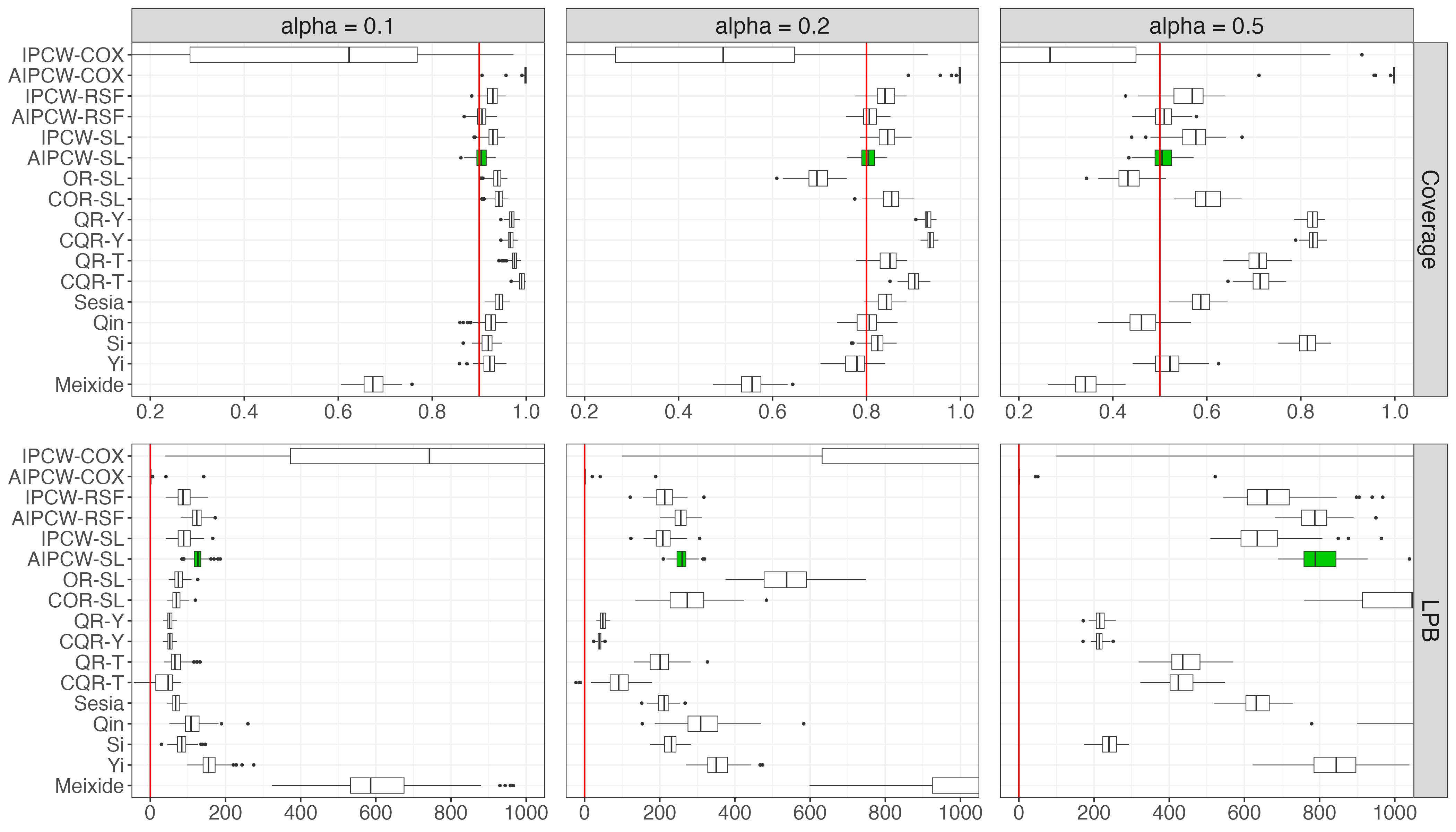}
    \caption{Distribution of empirical marginal coverage rates (top row) and average estimated LPB (bottom row) across 100 simulated test datasets under Setting 4 for each method, with different miscoverage levels $\alpha$. The red vertical lines in the top row denote the nominal coverage levels of 90\%, 80\%, and 50\% (from left to right), while the red vertical lines in the bottom row panel marks the lower limit of the support of the time-to-event outcome. The recommended method AIPCW-SL is highlighted in green.}
    \label{fig:boxplot_alpha_lognorm}
\end{figure}

For Settings 1 and 4, we repeated the simulations at target miscoverage levels $\alpha=0.2$ and $\alpha=0.5$; the corresponding empirical coverage and average LPB boxplots are shown in Figures~\ref{fig:boxplot_alpha} and~\ref{fig:boxplot_alpha_lognorm}. In Setting 1, when $\alpha=0.2$, the behavior is essentially unchanged from $\alpha=0.1$. When $\alpha=0.5$, IPCW/AIPCW, the naive baselines (QR-Y, CQR-Y, QR-T, CQR-T), and most competing methods show mild over-coverage. This is consistent with the fact that Setting 1 is a simple, low-dimensional exponential model with independent censoring, so most procedures can estimate the relevant nuisance functions reasonably well; once the target coverage is reduced to 50\%, even a modest downward bias in the estimated LPB translates into coverage above the nominal level. In other words, at this less stringent target, methods that are slightly conservative produce visibly higher-than-nominal coverage. The method of \cite{si2025training} is especially conservative here, with median coverage 0.79, and correspondingly very small LPBs. By contrast, the outcome-regression methods are the only procedures that maintain nearly exact calibration across nominal levels, which is unsurprising because in this setting the event-time model is simple and can be well learned by a flexible estimator such as SL.

In Setting 4, the IPCW/AIPCW procedures have similar behavior across all target coverage levels. This stability is consistent with the fact that these methods calibrate coverage directly through censoring-weighted estimating equations, so changing the nominal level mainly shifts the calibrated quantile level without fundamentally altering the estimation problem. In particular, with sufficiently flexible nuisance learners, the weighting-based correction continues to account for the heavy, covariate-dependent censoring even in this high-dimensional non-proportional hazards setting. In contrast, OR/COR and the naive methods become more conservative as $\alpha$ increases, since misspecification or estimation error in the conditional survival curve propagates directly into the bound, leading to LPBs that are too low and hence over-coverage. For the naive methods, the problem is more pronounced because they ignore censoring altogether or use only uncensored observations, which induces a downward bias in the estimated quantiles and thus increasingly conservative LPBs. The performance of competing methods also gets worse as $\alpha$ increases, suggesting that the combination of heavier censoring and a more difficult nuisance-learning problem makes calibration at lower nominal coverage more challenging. Again, \cite{si2025training} is notably conservative at $\alpha=0.5$.

The pronounced over-coverage of \cite{si2025training} at $\alpha=0.5$ in both simulation settings appears to be a finite-sample consequence of its conservative calibration rule. This effect becomes much more visible at $\alpha=0.5$, where the target corresponds to a more central part of the survival distribution and a downward shift in the LPB leads to a larger increase in empirical coverage. This phenomenon is likely related to the APAC guarantee targeted by the method, whose calibration is conservative by construction. In contrast, our AIPCW method targets marginal coverage directly and therefore does not exhibit this finite-sample conservativeness, remaining much closer to nominal across target levels.

\begin{figure}[t]
    \centering
    \includegraphics[width=\textwidth]{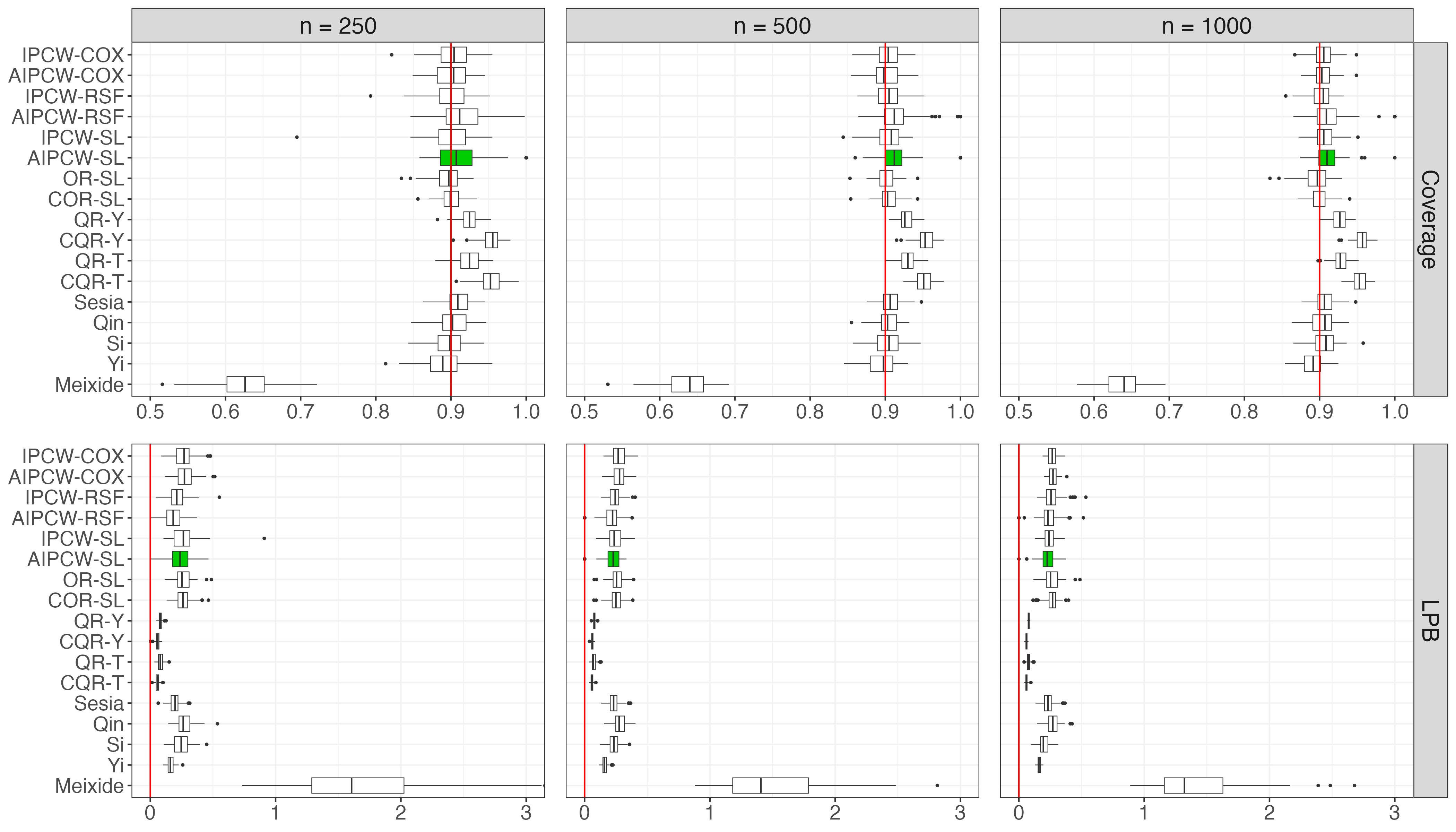}
    \caption{Distribution of empirical marginal coverage rates (top row) and average estimated LPB (bottom row) across 100 simulated test datasets under Setting 1 for each method, with different training and calibration sample sizes (250/250, 500/500, and 1000/1000). The red vertical lines in the top row denote the 90\% target coverage, while the red vertical lines in the bottom row panel marks the lower limit of the support of the time-to-event outcome. The recommended method AIPCW-SL is highlighted in green.}
    \label{fig:coverage_boxplot_sizes}
\end{figure}

\begin{figure}[h!]
    \centering
    \includegraphics[width=\textwidth]{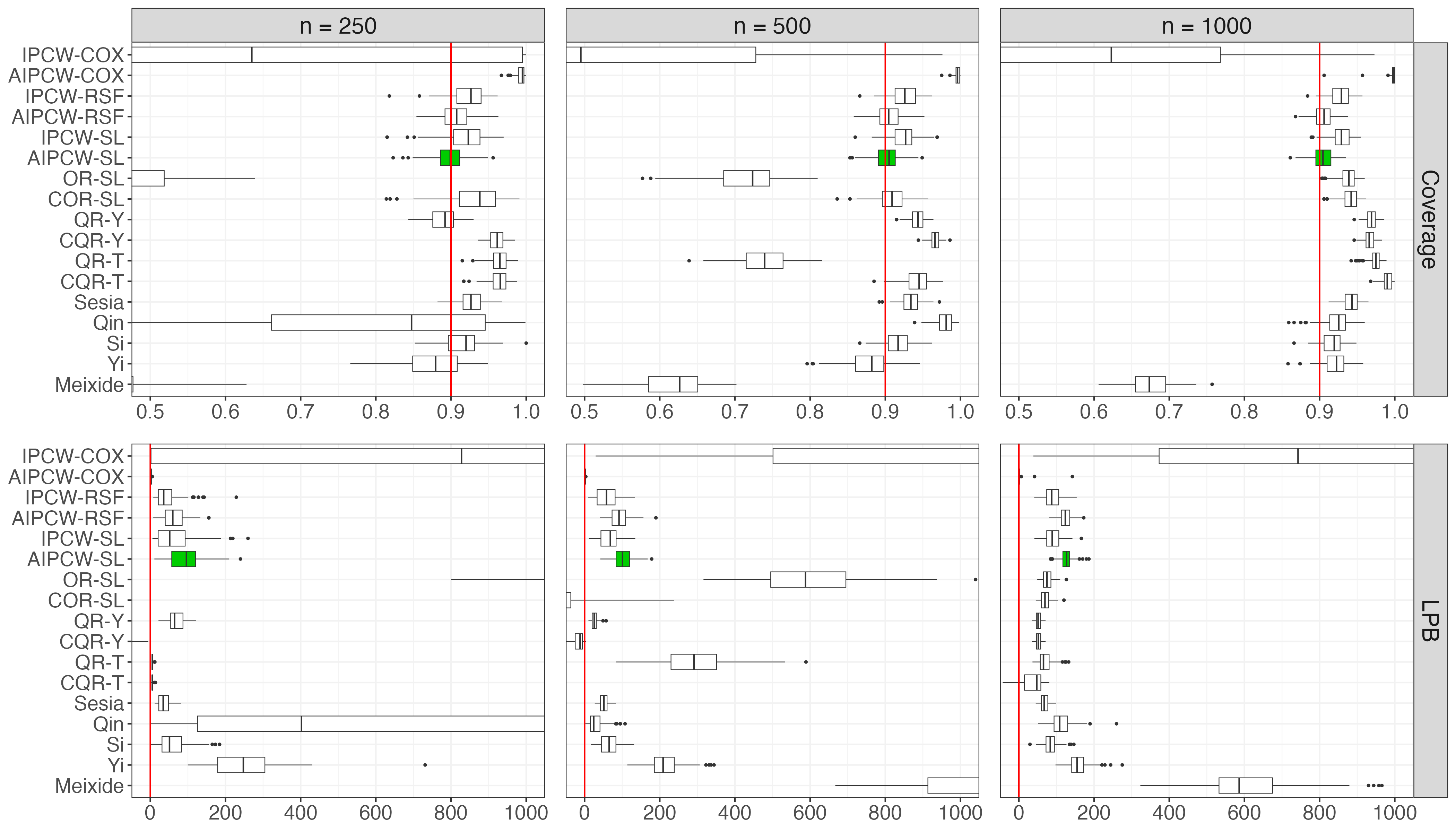}
    \caption{Distribution of empirical marginal coverage rates (top row) and average estimated LPB (bottom row) across 100 simulated test datasets under Setting 4 for each method, with different training and calibration sample sizes (250/250, 500/500, and 1000/1000). The red vertical lines in the top row denote the 90\% target coverage, while the red vertical lines in the bottom row panel marks the lower limit of the support of the time-to-event outcome. The recommended method AIPCW-SL is highlighted in green.}
    \label{fig:coverage_boxplot_sizes_lognorm}
\end{figure}

We also perform simulations with different training sample sizes: in the baseline scenario, we use 1000 training and 1000 calibration samples, while additional experiments are run with 500/500 and 250/250 training and calibration sample sizes, keeping the test set size fixed at 1000 for all experiments. The resulting empirical coverage and average LPB boxplots are displayed in Figure~\ref{fig:coverage_boxplot_sizes} for Setting 1, and Figure~\ref{fig:coverage_boxplot_sizes_lognorm} for Setting 4. %As the sample size decreases, the empirical coverage distribution becomes wider, reflecting increased variability. 
As expected, the empirical coverage distributions become wider as the sample size decreases, reflecting greater finite-sample variability. This behavior is unsurprising, since our theoretical guarantees are asymptotic and are therefore expected to hold more accurately at larger sample sizes.

In Setting 1, the overall ranking of the methods remains unchanged across sample sizes. The IPCW/AIPCW procedures remain close to the nominal level, with only an increase in dispersion as the training and calibration sets become smaller. The OR/COR methods continue to perform well, which is expected since the data-generating model is simple and can still be learned accurately even with fewer observations. The naive methods remain conservative across all sample sizes, suggesting that their behavior is driven primarily by their failure to properly account for censoring rather than by finite-sample instability. Among the competing methods, the patterns are also fairly stable: reducing the sample size mainly increases variability, rather than fundamentally changing their calibration properties.


In Setting 4, the effect of sample size is much more pronounced. This setting combines heavy covariate-dependent censoring with a more complex, high-dimensional, non-proportional hazards structure, so smaller training samples make nuisance estimation substantially harder. As a result, several methods show both increased variability and more noticeable deviations from nominal coverage when the sample size is reduced. The outcome-regression procedures are particularly sensitive to this change, especially the OR method, since any error in estimating the conditional survival function propagates directly into the LPB. The naive methods are conservative at all sample sizes, again because they fail to properly account for censoring. The competing methods also perform worse as the sample size decreases, especially \cite{qin2025conformal}. This is plausible given the structure of the method: it relies on a Cox working model, Kaplan–Meier censoring estimation, and bootstrap calibration, and in Setting 4, this strategy is more sensitive to smaller sample sizes. AIPCW-SL is stable across sample sizes and stays close to the nominal level, which is consistent with the fact that it directly adjusts for censoring and benefits from the augmented doubly robust correction. 

Overall, these additional experiments varying the target coverage level and the sample size further confirm the stability of the AIPCW-SL approach, which yields the most stable trade-off between calibration and LPB informativeness, especially in more challenging settings, with heavier censoring and when nuisance estimation is more difficult.

\begin{figure}[t]
    \centering
    \includegraphics[width=\textwidth]{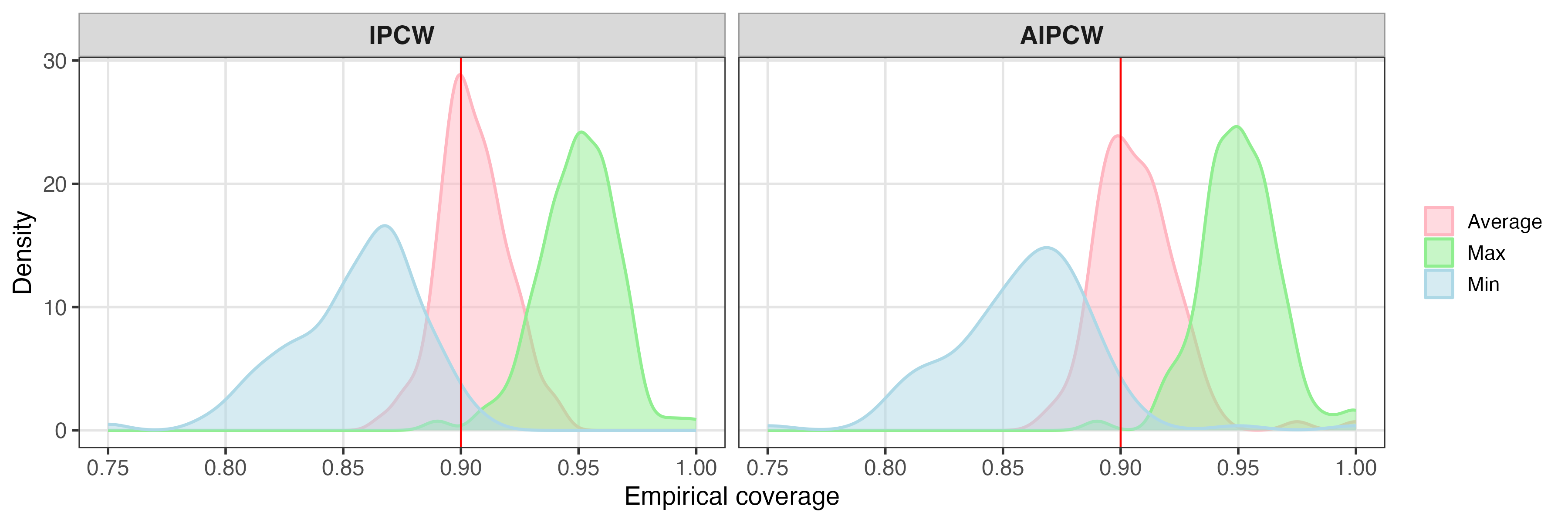}
    \caption{Distributions of minimum, mean, and maximum empirical conditional coverage across 10 test covariate values over 100 repetitions in Setting 1.
    }
    \label{fig:conditional_boxplot}
\end{figure}
\begin{figure}[t]
    \centering
    \includegraphics[width=\textwidth]{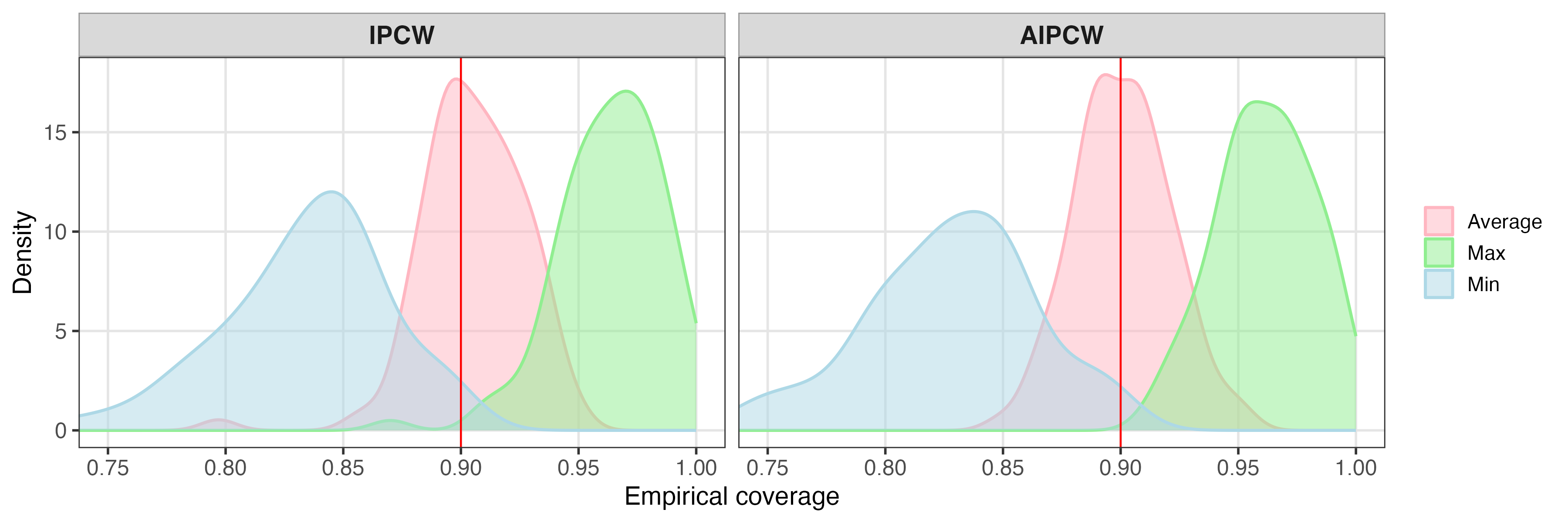}
    \caption{Distributions of minimum, mean, and maximum empirical conditional coverage across 10 test covariate values over 100 repetitions in Setting 6.
    }
    \label{fig:conditional_boxplot_set6}
\end{figure}

Finally, although the primary objective of this paper is marginal coverage, we conduct a simple conditional coverage simulation under Settings 1 and 6 using IPCW/AIPCW-SL. The training, calibration and test samples consist of 1000 observations each, as in the marginal coverage experiments. For the test set, we sample 10 values of $X$ and, for each $X$, generate 100 realizations of $T \mid X$. We then compute the average conditional coverage for each $X$. This experiment is repeated 100 times, and for each repetition we record the minimum, mean, and maximum empirical coverage across the 10 sampled $X$’s. The resulting distributions of these quantities are displayed in Figure~\ref{fig:conditional_boxplot} for Setting 1 and Figure~\ref{fig:conditional_boxplot_set6} for Setting 6. In both cases, the minimum and maximum conditional coverage distributions remain reasonably close to the target level 90\%, suggesting stable performance across different covariate values. The mean remains very close to the target, as expected from averaging over the covariate distribution. As expected, the behavior is slightly more variable in Setting 6 than in Setting 1, since Setting 6 is substantially more challenging due to its high-dimensional, heteroskedastic event-time model and covariate-dependent censoring; nevertheless, the results remain broadly stable across sampled covariate values in both settings.

\section{Additional real data results}\label{App:data}
Table~\ref{tab:SL_data} shows the estimated coefficients of the candidate models used within the SL algorithm.
\begin{table}[h!]
    \centering
    \caption{Aggregated average of the model coefficients over the 100 splits.}
    \begin{tabular}{lcc}
        \toprule
        Model & $T\mid X$ & $C\mid X$ \\
        \midrule
        km        & 0.052  & 0.000  \\
        coxph     & 0.014  & 0.320  \\
        expreg    & 0.217  & 0.000  \\
        weibreg   & 0.025  & 0.001  \\
        loglogreg & 0.133  & 0.155  \\
        gam       & 0.020  & 0.018  \\
        rfsrc     & 0.540  & 0.506  \\
        \bottomrule
    \end{tabular}
    \label{tab:SL_data}
\end{table}

\end{document}